# Cost-effectiveness Analysis of Antiepidemic Policies and Global Situation Assessment of COVID-19

**Authors:** Liyan Xu[1]*, Hongmou Zhang[2]*, Yuqiao Deng[1], Keli Wang[3], Fu Li[1], Qing Lu[4], Jie Yin[5], Qian Di[6], Tao Liu[4], Hang Yin[1], Zijiao Zhang[1], Qingyang Du[1], Hongbin Yu[1], Aihan Liu[1], Hezhishi Jiang[1], Jing Guo[1], Xiumei Yuan[1], Yun Zhang[1], Liu Liu[7], Yu Liu[3]†

**Affiliations:**

[1]College of Architecture and Landscape Architecture, Peking University, Beijing 100871, China.

[2]Future Urban Mobility IRG, Singapore–MIT Alliance for Research and Technology, Singapore 138602.

[3]School of Earth and Space Sciences, Peking University, Beijing, 100871, China.

[4]College of Urban and Environmental Sciences, Peking University, Beijing 100871, China.

[5]Department of Environmental Health, Harvard T.H. Chan School of Public Health, Boston, MA 02115, USA.

[6]Research Center for Public Health, School of Medicine, Tsinghua University, Beijing 100084, China

[7]CitoryTech, Shanghai 200063, China.

*These authors contributed equally to this work.

†Corresponding author. Email: liuyu@urban.pku.edu.cn.

**Abstract:** With a two-layer contact-dispersion model and data in China, we analyze the cost-effectiveness of three types of antiepidemic measures for COVID-19: regular epidemiological control, local social interaction control, and inter-city travel restriction. We find that: 1) inter-city travel restriction has minimal or even negative effect compared to the other two at the national level; 2) the time of reaching turning point is independent of the current number of cases, and only related to the enforcement stringency of epidemiological control and social interaction control measures; 3) strong enforcement at the early stage is the only opportunity to maximize both antiepidemic effectiveness and cost-effectiveness; 4) mediocre stringency of social interaction measures is the worst choice. Subsequently, we cluster countries/regions into four groups based on their control measures and provide situation assessment and policy suggestions for each group.

Coronavirus disease 2019 (COVID-19) has been recognized as a pandemic by the World Health Organization in March 2020 (*1*). Nonetheless, on March 18, mainland China observed their first day with zero increase of local cases since the outbreak (*2*). This indicates that, aside from the imported cases, the pandemic has been locally close to the end. Many experiences and lessons can be shared by the rest of the world from the trajectories of the outbreak and the control strategies in mainland China. One question that is of particular importance is the cost-effectiveness of different antiepidemic measures. Drawn from the Chinese experiences and lessons, three categories of measures have been implemented: a) "regular" epidemiological control and prevention measures, including identification of infected cases, tracing their close contacts, and quarantines for both; b) *In-city* activity restrictions, including work-from-home, shutdown of schools and public spaces, cancellation of events, and lock-down of residential neighborhoods; c) *Inter-city* travel restrictions, including temperature screening at all transportation terminals, cancellation of flights and trains, and eventually travel bans from/to certain cities. Specifically, (b) and (c) are considered "irregular," which contain the spread of disease in an aggressive manner *through the suppression of all possible social interactions*. However, these measures lead to enormous economic loss, which may mean higher unemployment rates, and shortage of food, medical services, and other necessities. Those chain effects may also endanger the lives of certain social groups. Fundamentally this is a trolley dilemma, and the life and health of human beings can hardly be evaluated using monetary values. Nevertheless, a comprehensive understanding of the cost of each measure—including the opportunity cost of the shutdown economic activities—and the effectiveness of lifesaving can still help policy makers to compare different antiepidemic strategies in a more operable way. Further, with these insights, we can then perform a cross-sectional assessment of the situation of the global antiepidemic campaign regarding the aforementioned measures in different countries/regions, such that typical policies can be clustered and prescriptive policy suggestions can be provided accordingly, which also echoes the argument in (*3*).

In prior research, scholars have used deductive models to find that the *timing* of lock-down, including both inter-city travel restrictions and social distancing, significantly changes the size of infected population and the spatial extent of spread (*4*–*7*). Nonetheless, all prior research only focused on specific instances of policies without discussing generalizable impacts of different measures and lacked a cost-effectiveness assessment of each measure (*8*–*10*). In this article, based on the transmission pattern of COVID-19 in China, we build a two-layer contact-spread model (Fig. 1) to recover the whole spatio-temporal transmission process, especially the early-stage numbers and distribution of cases at the prefecture level (see SM 1, 2, and 4). With this model as a generalizable baseline, we conducted a comprehensive sensitivity analysis for all antiepidemic measures (Fig. 2) and concluded the following assessment on the *effectiveness* of the measures, in terms of number of infected cases, and number of infected cities.

1. Containing daily social interaction, parameterized in the model as $\kappa_I$, $\kappa_E$ for the infected and the exposed populations, is *the most effective measure* for controlling both the number of infected cases and the spatial extent of the spread. More specifically, we find that a) the controls of $\kappa_I$ and $\kappa_E$ show comparable and substitutable effects in containing the spread of disease, with different elasticity in different stages of the epidemic. Controlling $\kappa_E$ is more effective when the number of cases is small (e.g., fewer than 10 in a city), and $\kappa_I$ is more effective when the number of cases is sufficiently large. The implication is that at the early stage of transmission, comprehensive epidemic surveys and contact tracing, alongside with strict quarantine and social distancing, should be used to prioritize the reduction of the social interaction levels of the exposed population ($\kappa_E$), while after the number of cases has increased to a sufficiently high level, comprehensive testing and identification of all infected cases should be prioritized in order to reduce the social interaction levels of the infected population ($\kappa_I$). b) Stricter control of social interaction for both $\kappa_I$ and $\kappa_E$ have *diminishing returns in reducing the number of cases, but increasing returns in limiting the spatial extent of spread*. For example, in one month of simulation, when daily social interaction drops from 100% to 90% of the normal level, the number of infected cases drops by 10%–25% (depending on the stage of epidemic), while the number of cities with infected cases drops only by 0–1%; when daily social interaction drops from 10% to 0, however, the number of cases drops only by 0.05%–0.5% for infected cases versus 2%–35% for cities with infected cases. An exception is at the ending stage of the epidemic, when controlling social interaction has increasing returns in both effects. c) The effects of the epidemiological and social interaction control measures are *monotonic* for the reduction of infected cases and the spatial extent of spread: the stricter they are enforced, the lower number of infected cases and the narrower spatial extent of spread can be observed.

2. The time of reaching turning point is *independent* of the current number of infected cases but is only related to the stringency of epidemiological and social interaction control measures, i.e., the relative change of $\kappa_I$, $\kappa_E$. When the two parameters are 1/4–1/3 of the normal everyday values, the turning point comes in two weeks and the clearance of cases happens in two to three months; when $\kappa_I$, $\kappa_E$ are larger than 1/2–2/3 of the normal values, the turning point will never come, i.e., the peak value of case numbers will remain the same as if there are no such measures, but they only delay the time of peak.

3. Except at the early stage and the ending stage, inter-city travel restriction has only minimal effect on both the reduction of infected cases and control of disease spread in a city network. Overall, compared with in-city epidemiological and social interaction control measures, the contribution of inter-city travel restrictions to the reduction of the number of infected cases and the spatial spread of disease is much smaller—lower by two orders of magnitudes. When the number of cases is sufficiently large, inter-city travel restriction even exacerbates the situation since it limits the social interaction of infected cases, and "condenses" $R_0$ locally (*11*). This finding is consistent with that in prior research (*12*, *13*). Therefore, to national or regional governments who manage a city network, and to international antiepidemic collaboration, travel restriction should only be regarded as an auxiliary measure at the beginning and ending stage of the spread to protect cities which have not been infected at all, or only with a sufficiently small number of cases. In the latter case epidemiological and social interaction control measures should also be implemented simultaneously to get the health care system and other prevention measures prepared.

The simulation-based formal analysis above is consistent with the empirical evidence in China. The lockdown of Wuhan, and nationwide strict enforcement of epidemiological control and social distancing policies around January 23, including the cancellation of all Chinese New Near gatherings mark the key move of the antiepidemic campaign. At that time point, all other cities in China were at early stages of the epidemic, which guaranteed the effectiveness of the Wuhan travel ban. In addition, with aggressive social interaction control in all cities the turning point of the number of cases arrived in two weeks outside of Wuhan. In Wuhan, the key move was the functioning of 16 *fāngcāng* hospitals (mobile cabin hospitals) in early February which enabled citywide comprehensive quarantine of the infected population (*14*). This measure reduced the social interaction of infected cases to almost *zero*, and together with strict social distancing they effectively reversed the trend of spread after two weeks. In terms of inter-city travel restrictions, since they were during the Chinese New Year and the extended holidays, and overlapped with social distancing measures, the net effect could not be easily isolated empirically. Nonetheless, since mid-February the economy had re-opened. By the end of March, the inter-city migration in southern and eastern Chinese cities had recovered to the same level as in previous years (*15*), but most cities still observed almost zero increase of infected cases. This further supports that the effectiveness of travel restriction is very limited for the cities with small numbers of cases. Another evidence to support this point is the 300,000 people who left Wuhan right before the lockdown night (*16*). This "escaped" population did not significantly change the effectiveness of the national antiepidemic effort.

Further analysis on the cost-effectiveness of the measures shows more irregularity and non-linearity, leading to more nuanced relationships (detailed in SM 5). Here we summarize the most critical general patterns as follows:

1. The measures which can achieve both high antiepidemic effectiveness (low number of cases and narrow spatial spread) and high cost-effectiveness (smaller loss of economic outputs) only exist at the early stage of transmission. At the early stage, if epidemiological and social interaction control measures can be strictly enforced (sufficiently low $\kappa_I$ and $\kappa_E$), it is possible to keep the spread at a low level, with a loss of economic outputs only up to 4%. The intuition is as follows: based on the assumptions of this article, the early-stage measures only include comprehensive testing, close contact tracing, and quarantine, but do not include indiscriminate restrictions of in-city social interaction and inter-city travel, which incurs high costs. The policy implication is straightforward: for early-stage cities and regions, it is critical to practice epidemiological control interventions, but not to necessarily mobilize the whole society into social interaction reduction. This finding is consistent with the suggestions in (*6*).

2. Except for the early stage, it is *impossible to simultaneously achieve both high antiepidemic effectiveness and high cost-effectiveness*. Except for a few "plateaus," the effectiveness of epidemiological and social interaction control measures monotonically increases with the stringency of control measures. However, the cost and cost-effectiveness functions are non-monotonic and there usually exists more than one peak (see details in SM 5), which in most cases do not coincide with the effectiveness peak. Typically, the costs are the lowest when the control measures are at sufficiently low or sufficiently high levels. While the latter case has been explained in the last point, the sufficiently low control measure scenario basically leaves the whole population to be infected.

Therefore, the tradeoff between sufficiently low and sufficiently high levels of control measures depend on many technological factors, including the short-term and long-term capacity of healthcare systems, long-term uncertainty of virus mutation, and development of vaccines, as well as many non-technological factors, including the risk averse attitudes for the short term and the long term, the mental discounting between short-term and long-term tradeoffs, and the fundamental value judgement on the "value of lives," the discussion of which are beyond the scope of this article, and will be left for discussion at the end of this article.

3. Lastly, although it is difficult to choose the optimal control strategy, the worst choice is explicit: *mediocre control of social interaction*, e.g., social distancing with leakage. This choice still incurs 20–60% loss of economic outputs, but only achieves 30–40% reduction in the number of cases, an extent which is insufficient to overturn the epidemic curve. Except for moderately delaying the spread of disease which may be taken advantage of to get the healthcare system prepared, this strategy is the worst choice in all other dimensions.

With the formal results above, we can now perform a cross-sectional assessment of the global situation of the antiepidemic campaign from a transmission-prevention policy perspective. Among the three types of measures (epidemiological control measures, social distancing, and travel restriction), we disregard the travel restriction measure as our results clearly show that it is ineffective for most countries/regions under the current situation (we will discuss the exceptions later). Rather, we use two datasets (*17*, *18*) which codified the antiepidemic measures chosen by countries/regions as of April 7 (due to the lack of testing data, we use Wuhan as a proxy for mainland China), and for the countries/regions analyze the relationship of their two stringency indices, $\kappa_i$ and $\kappa_e$, i.e., the activity levels of the infected and exposed populations, and the respective effectiveness on the reduction of infected cases. Based on the stringency of the two dimensions of antiepidemic measures, we can divide all countries/regions into three groups (Fig. 3), each with a different antiepidemic "strategy": *elimination*, *control*, and *delay*. More than 100 countries/regions are not included because of the lack of data. We will also discuss the implications of this fact.

1. The "elimination" group: This group (up right corner of Fig. 3) consists of only a few countries/regions, including mainland China (represented by Wuhan), Hong Kong SAR, Vietnam, UAE, Bahrain, etc., all with $R_0 \ll 1$, such that the epidemic could be expected to dwarf within a reasonably short time period. Mainland China is the most prominent example of this group, where aggressive measures have been taken on both dimensions to reduce the activity of the infected population as well as the exposed population. The measures include effective epidemiological control interventions, such as comprehensive testing and close contact tracing, and also aggressive social distancing measures, such as shutdown of schools, workplaces, and public transport, cancellation of events, and mass disease control education. These measures incur 40%–90% loss of economic outcome in a month, and the loss accumulates as the epidemic is not completely "eliminated". Obviously, the underlying value judgment of the elimination strategy is an overwhelmingly high weight on health and lives over any cost-control or cost-effectiveness reckoning.

Although the treasuring for lives is always respectable, long-lasting economic tightening also constitutes a threat to society, especially to the disadvantaged social groups. Due to the existence of asymptomatic carriers, false-negative test results, and international imports of cases, a complete elimination of the epidemic is extremely difficult. Thus, if the aim is to literally eliminate all cases, the economic losses are highly likely to accumulate to an unbearable level. Therefore, we suggest that countries/regions which have followed the elimination strategy consider turning to the "control" strategy (elaborated below) to avoid excess economic losses on the condition that the active number of infected cases has been reduced to a sufficiently low level. We also suggest that these countries/regions keep the travel restriction measures—the most effective measure at this stage of the epidemic indicated by our simulation results.

2. The "control" group: This group includes South Korea, Singapore, Qatar, Norway, Slovenia, Russia, and New Zealand, etc., all with $R_0 < 1$, but still not sufficiently small, such that the epidemic can be reduced to a lower level (but not eliminated), depending on the stringency of intervention measures. The Singapore in February was the most prominent example within this group, where antiepidemic measures have been mild enough not to affect everyday life by aggressive social distancing. Through regular epidemiological control practices, they were managed to maintain a daily increase of infected cases fewer than 10, and only suffered 0.5%–4% loss of economic outcome in a month. The control strategy requires a highly capable epidemic control system. Given the aforementioned long-term uncertainties, even with such a capable system, the strategy is still a tightrope-walking game with the risk of abrupt system overload by accidentally untracked surges of infection, which, unfortunately, appears to be the case in Singapore in early April. Under such circumstances, a timely turn to the "elimination" strategy may be necessary.

3. The "delay" group: All other countries/regions in Fig. 3 belong to the third group, which appears to follow the "delay" strategy, with $R_0 > 1$, such that the epidemic will continue to grow. This is often referred to as the "flatten the curve" strategy, which aims not to reduce the epidemic to an as-low-as-possible level within a short period of time, but only to delay its growth through mediocre epidemiological control and social distancing measures. Our results show that this is usually the worst scenario in terms of cost-effectiveness. A country/region may opt to this strategy because their tradeoff between short-term certainty (economic loss avoidance) and long-term uncertainty (possible disappearance of the epidemic in the summer, development of vaccines, etc.) leans towards the former. Unless they have strong evidence to justify the tradeoff, we strongly suggest they reconsider. Moreover, our results show possible directions to improve—enhancing the social interaction control for the infected population through more comprehensive testing or enhancing the social interaction for the exposed population through stricter social distancing measures, whichever sees fit based on the location of the country/region on Fig. 3.

4. Rest of the world: More than 100 countries/regions do not appear in Fig. 3 due to the lack of data, most of which are third-world countries/regions. Although little information is available to us about the situations in these places, we conjecture that they may at this moment be pursuing cost-effectiveness of their antiepidemic interventions because of their limited availability of resources, which we call the "worth every penny" strategy. As our results show that the most cost-effective measures are usually neither the most effective one (actually they are usually very *ineffective*), nor the least costive ones, the "worth every penny" strategy is not a good option either. If a country/region opts to this scenario solely because of the lack of resources, it should be viewed as a humanitarian disaster, and we call for international aid in this situation.

At the end, we acknowledge the extreme difficulty of even trying to lay out the comparison between human lives and economic activities, or the tradeoffs of lives between different social groups. We believe that the ethical discussion should be open to the whole society and hope that this article can contribute to the discussion.

**Fig. 1. Structure diagram of the contact-dispersion model.** The model is consisted of an in-city layer of SEIR model (*19*) and a network transmission layer based on inter-city migration. Through inter-city travel, the numbers of exposed and infected populations are adjusted daily. The model is calibrated using the migration data and the number of reported cases in China. See Materials and Methods for model specifications.

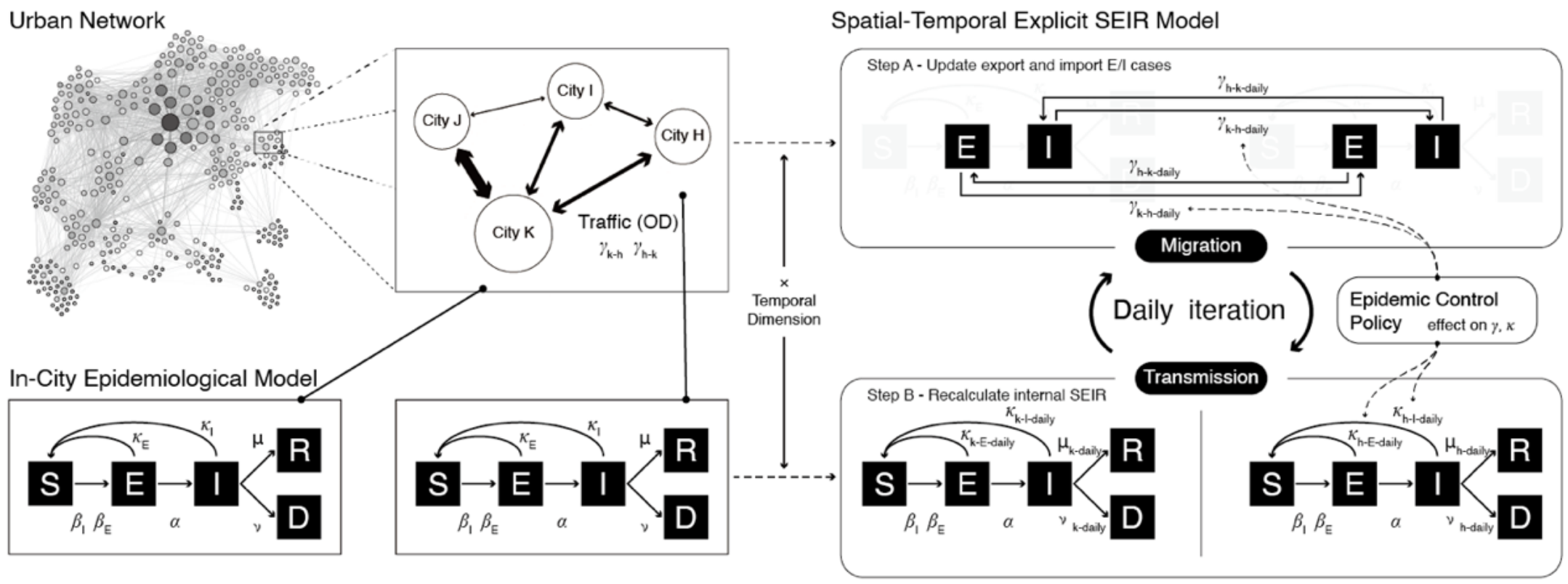

**Fig. 2. Gradients of the cost, two metrics of effectiveness, and the two respective cost-effectiveness functions with respect to social interaction control indicators ($\kappa_I$, $\kappa_E$) and the inter-city travel level indicator at different stages of the epidemic.** (**a**) Gradient of the first effectiveness function (with the number of infected cases as the metric for effectiveness) at the peak/inflecting stage, displaying decreasing margins, as well as the reverse of the effect of the inter-city travel level as $\kappa_I$ and $\kappa_E$ decreases. (**b**) Gradient of the second effectiveness function (with the number of cities with infected cases as the metric for effectiveness) at the accelerating stage, displaying firstly increasing and then decreasing margins as $\kappa_I$ and $\kappa_E$ decreases. (**c**) Gradient of the second effectiveness function (with the number of cities with infected cases as the metric for effectiveness) at the ending stage, displaying monotonically decreasing margins as $\kappa_I$ and $\kappa_E$ decreases. See definition of different stages and the calculation of gradient in Materials and Methods. (**d**) Gradient of the cost function (labor-hour loss) at the accelerating stage, displaying varying margins and a "ridge" of the cost. (**e**) Gradient of the first cost-effectiveness function at the early stage, displaying varying margins, and changing directions of the contribution of the inter-city travel level coefficient. (**f**) Gradient of the second cost-effectiveness function at the early stage, displaying increasing margins as $\kappa_I$ and $\kappa_E$ decreases and relatively insignificant contribution of inter-city travel level.

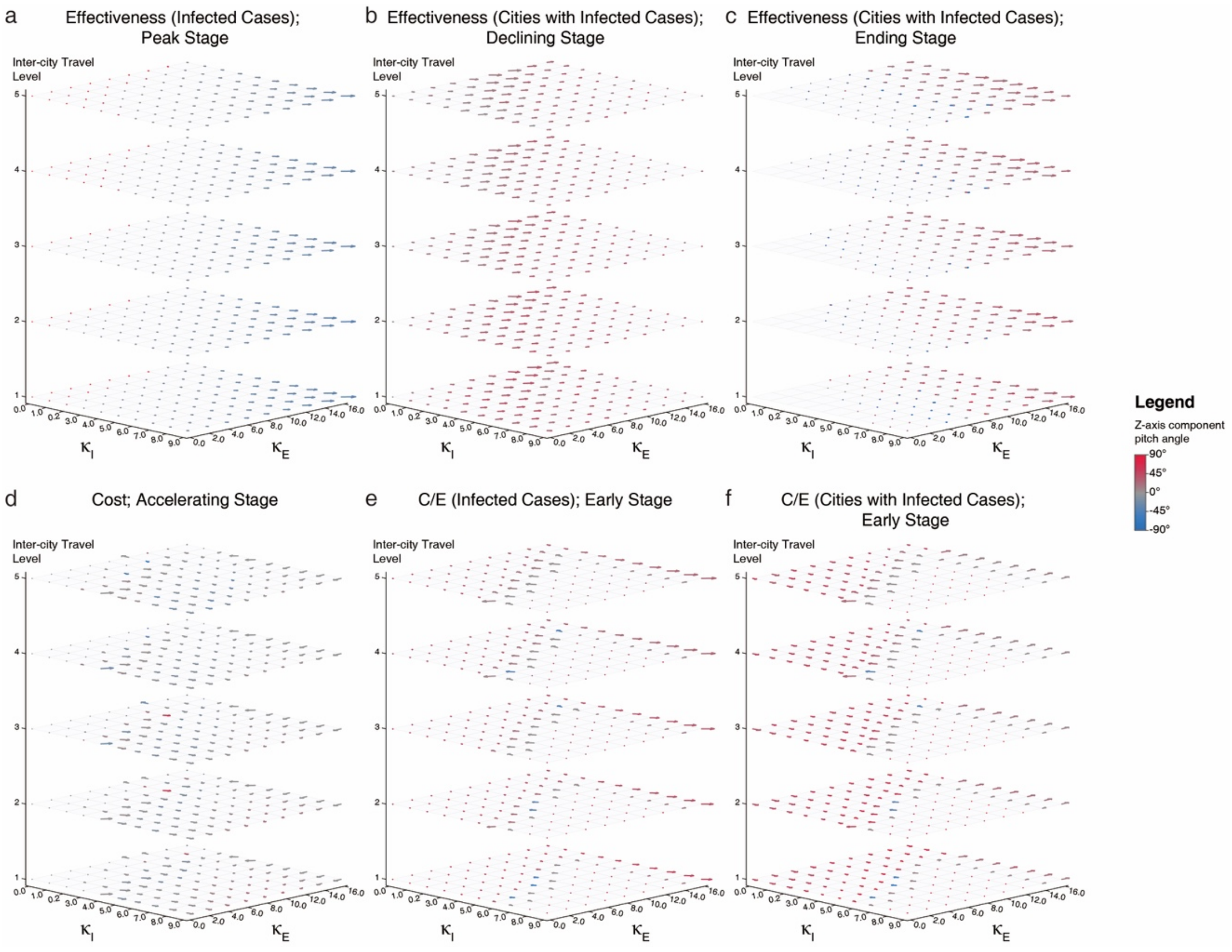

**Fig. 3. Groups of antiepidemic policy based on stringency of social interaction measures.** Each dot represents a country/region. The sizes of the dots indicate the number of infected cases on April 7, 2020.

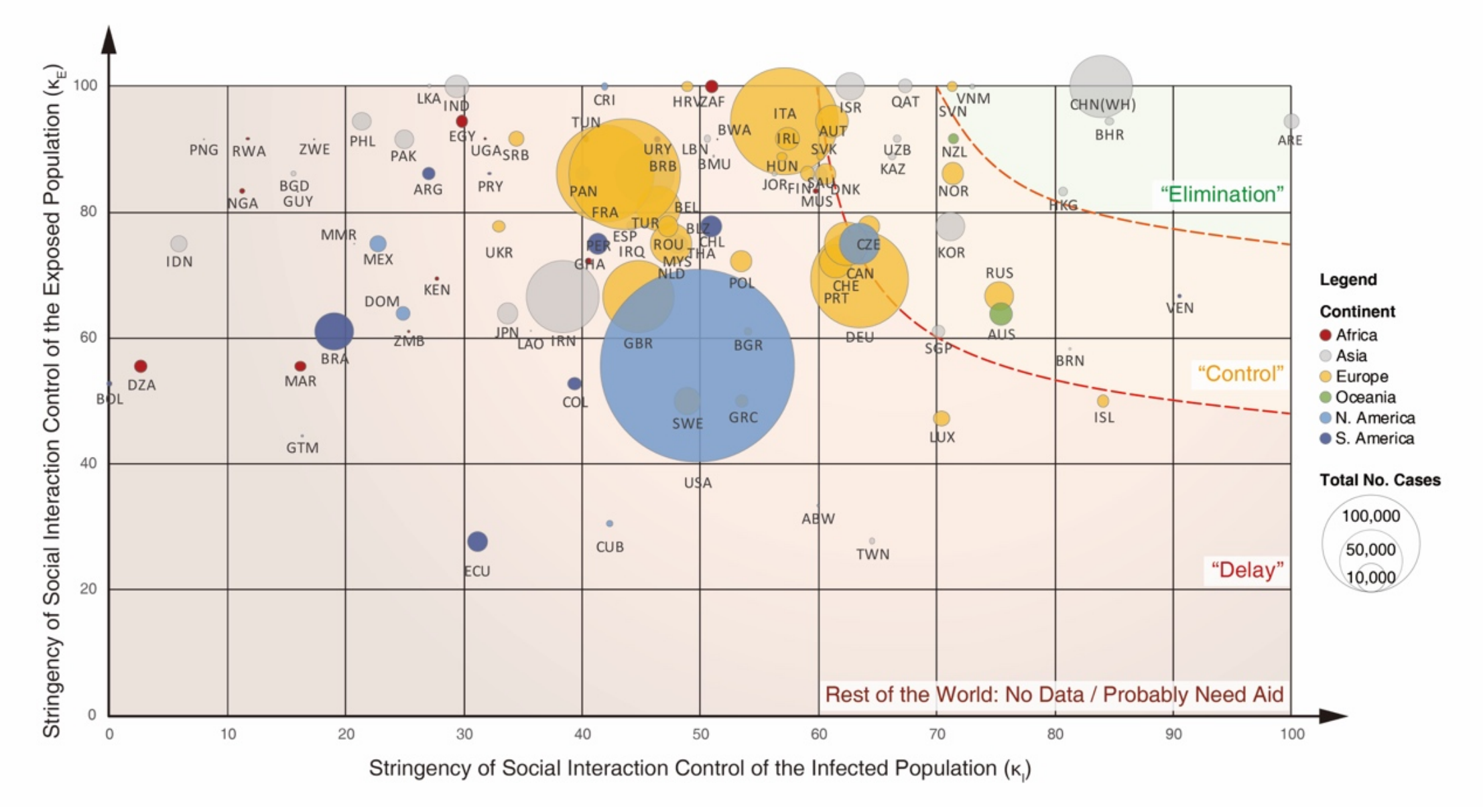

Supplementary Materials for

# Cost-effectiveness Analysis of Antiepidemic Policies and Global Situation Assessment of COVID-19

**Authors:** Liyan Xu*, Hongmou Zhang*, Yuqiao Deng, Keli Wang, Fu Li, Qing Lu, Jie Yin, Qian Di, Tao Liu, Hang Yin, Zijiao Zhang, Qingyang Du, Hongbin Yu, Aihan Liu, Hezhishi Jiang, Jing Guo, Xiumei Yuan, Yun Zhang, Liu Liu, Yu Liu†

*These authors contributed equally to this work.

†Corresponding author. Email: liuyu@urban.pku.edu.cn.

**This PDF file includes:**

Materials and Methods
Figs. S1 to S7
Tables S1 to S11

## Materials and Methods

### 1. A Spatial-Temporal Explicit SEIR (STEx-SEIR) Model

Classical SEIR model characterizes the dynamics of the susceptible, exposed, infected, and removed population, assuming a certain population size and transmission rate. In closer scrutiny, the fixed transmission rate assumption requires contact-based transmission process (*19*). From the perspective of spatial interaction, this requirement is contradictory to the prior assumption of fixed population. Due to the high mobility of modern society, fixed population requires a large spatial scale of the model, e.g., the national scale. Moreover, contact-based assumption is valid only at the scale of human *daily* activities—such scale should not exceed the urban scale to incorporate multiday intercity travels.

At a larger spatial scale, human activity is featured by the "return–explorer" dichotomy (*20*). The former spreads outbreaks through daily contact, while the latter "diffuses" the epidemic through long-distance travels. The order of magnitude of the intercity long-distance migration can be massive in China, and related studies have also discussed the effect of this population flow on the spread of COVID-19 outbreak (*8*, *9*, *21*, *22*). Thus, the COVID-19 outbreak is actually subject to a "contact-spread" two-step transmission process. Therefore, the classic SEIR model is insufficient to accurately describe the epidemic dynamics in the *urban network* and needs to be expanded.

Thereupon, we constructed a series of SEIR models for all cities, coupled with an intercity network. The node-edge structure of the model corresponds to the above-mentioned contact-spread process; and the required data (epidemic parameters of each city, and intercity spatial interaction) are relatively simple. Here, the choice of appropriate spatial resolution is critical. On the one hand, while prior research have adopted similar model design, they conducted the analysis mostly at the *provincial* resolution (*5*, *23*, *24*). According to the analysis above, the provincial level is not the best choice to reflect the intercity-spread mechanism. On the other hand, at too fine spatial scales, such as 1 km-grid, township, or census blocks, modeling results can hardly be verified due to the lack of data, and random errors can be large. Attempting to make a compromise between simplicity and completeness, interpretability and prediction accuracy, data spatial accuracy and verifiability (*25–27*), we took cities (administratively called *prefectures* in China) as the unit of analysis and built a model which we call the *Spatial-Temporal Explicit SEIR Model* (STEx-SEIR). This model can not only be used to analyze the number of infected cases, spatial range of infection, and dynamics of the COVID-19 outbreak, but can also be used to reveal the arrival time of the first case in each city. We calibrate the model with Chinese case statistics. Although it does not cover the entire globe, it is still informative to other countries since China has gone through the whole process of the epidemic.

#### 1.1 Model Assumptions

The STEx-SEIR model relies on the following assumptions: 1) Intra-city population is homogeneous, i.e., it is a "meta-population," which is a common practice in the absence of real demographic data; 2) Omitting the impact of foreign imports and exports of COVID-19 cases: the proportions of imported and exported cases during the COVID-19 outbreak in China are extremely small and can be omitted over the study period. 3) Virus remains unmuted: although a very large number of mutations have been recorded (*28*), the transmission patterns of the virus have not noticeably changed so far.

### 1.2 Model Specification

First step: update the immigrated and emigrated populations, the exposed population, and the infected population (assuming that the inputs and outputs of $N$, $E$, and $I$ across cities are completed instantly at the beginning of time $t$).

$$N_i(t) = N_i(t-1) - \Delta_{\text{Out}} N_i(t) + \Delta_{\text{In}} N_i(t)$$

$$\begin{cases} I_i(t) = I_i(t-1) - \Delta_{\text{Out}} N_i(t) \dfrac{I_i(t-1)}{N_i(t-1)} + \sum_j \left(\Delta_{\text{Out}} N_i(t) \dfrac{I_i(t-1)}{N_i(t-1)}\right) & t \le t_{\text{policy}} \\ I_i(t) = 0 & t > t_{\text{policy}} \end{cases}$$

$$E_i(t) = E_i(t-1) - \gamma \Delta_{\text{Out}} N_i(t) \frac{E_i(t-1) + I_i(t-1)}{\alpha N_i(t-1)} + \sum_j \left(\gamma \Delta_{\text{Out}} N_i(t) \frac{E_i(t-1) + I_i(t-1)}{\alpha N_i(t-1)}\right)$$

$$\begin{cases} \gamma = 1.5 & t \le t_{\text{control}} \\ \gamma = 1.1 & t_{\text{control}} < t \le t_{\text{policy}} \\ \gamma = 1 & t \ge t_{\text{policy}} \end{cases}$$

- $t_{\text{policy}}$: Lockdown on January 23;
- $t_{\text{control}}$: Measures taken on January 20; human-to-human transmission was confirmed on that day.
- $\gamma$: The proportion of the susceptible turning into the exposed status.

Second step: SEIR model in each city:

$$\frac{dS_i(t)}{dt} = -\kappa_I \beta_I \frac{I_i(t)}{N_i(t)} S(t) - \kappa_E \beta_E \frac{\theta_E E_i(t)}{N_i(t)} S(t)$$

$$\frac{dE_i(t)}{dt} = \kappa_I \beta_I \frac{I_i(t)}{N_i(t)} S(t) - \alpha E_i(t) + \kappa_E \beta_E \frac{\theta_E E_i(t)}{N_i(t)} S(t)$$

$$\frac{dI_i(t)}{dt} = \alpha E_i(t) - \mu I_i(t) - \nu I_i(t)$$

$$\frac{dR_i(t)}{dt} = \mu I_i(t)$$

$$\frac{dD_i(t)}{dt} = \nu I_i(t)$$

In the above equations: $S$ (unimmune and susceptible population) may turn into infected by contacting infected individuals; $E$ (exposed) is the population who are in the incubation period after being infected; $I$ (infected) is the population being symptomatic and infectious; $R$ (removed) is the population who died or have been cured—they can neither infect others nor be infected again. $N$: total population in the city, subject to $N = S + E + I + R$.

Other notations:

- $\beta_E$: Probability that a susceptible is infected and becomes an exposed person;
- $\beta_I$: Probability that a susceptible is infected by an exposed person;
- $\alpha$: Probability that an exposed person turns into infectious status;
- $\mu$: Probability of being cured;
- $v$: Probability that an infected person dies;
- $\theta_E$: Probability that an exposed person is infectious;
- $\kappa_I$: The number of susceptible persons who an infectious individual contact during each step of simulation;
- $\kappa_E$: The number of susceptible persons who an exposed individual contact during every step of simulation.

## 1.3 Model Calibration

### 1.3.1 Model Initialization

The case data used in this article come from the national and provincial health commissions in China (*29*). Wuhan is the city hit hardest by the epidemic, accounted for 61.87% of the total number of cases. The number of cases in Hubei province other than Wuhan accounted for another 22.03%. Therefore, Wuhan and Hubei Province have very different medical conditions, prevention and control measures from other cities and provinces in China. Thus, we assigned different initial parameters for Wuhan, other cities in Hubei, and cities in other provinces. The initial number of cases in Wuhan was set to be 1, and the initial numbers in other cities in Hubei and in other provinces was set to be 0.

According to the national report, the first confirmed infected case in Wuhan can be traced back to December 8, 2019. Thus, we chose December 8, 2019 as the starting point for the simulation. According to other studies (*30*), this date is as close as possible to the real starting point of the pandemic, and epidemiological data are only available from this point to calibrate model.

The endpoint of model calibration is February 26, when the epidemic in China was almost over and the model has converged by that time. We will simulate the spread of the COVID-19 epidemic from that date until the end of April as verification and prediction (by the time of writing this article).

### 1.3.2 Baseline Calibration

We used Chinese national, provincial, and municipal reports as the baseline to calibrate our model, so that the fitted number of cases would match with the surveyed data. On the one hand, the assumption, based on case statistics, is reliable. Reports are released separately by more than 300 cities nationwide and updated daily. Panel data from different provinces are inter-validated with other and are also supported by data from other sources (e.g., total mortality rate of the population, or mortality data of similar symptoms such as influenza, pneumonia, etc.). Thus, it is extremely unlikely that the data are fabricated *continuously without being noticed*. On the other hand, research shows that due to delayed diagnosis at the early stage of outbreak, the actual onset date of most confirmed cases is actually one week ahead of the report date (*30*, *31*). Considering the incubation period, we pushed the outbreak curve from the reported dates backwards by a few days, thus yielding the baseline for model calibration.

### 1.3.3 Parameter Values

The calculation and initial values of $N, \alpha, \beta_1, \beta_2, \gamma, \mu, v, \theta_E, \kappa_I, \kappa_E$, and $\Delta N$ were articulated in Section 4 of the Supplementary Material.

## 2. Effectiveness and Cost-effectiveness Analysis of Antiepidemic Measures

### 2.1 Control Measure Categorization

Among the three basic dimensions of epidemic control (control of the source of infection, cutting off transmission, and protection of the susceptible), we focus on the transmission cutting-off dimension. Specific measures belonging to this dimension can be further divided into three categories: 1) "regular" epidemiological control Measures; 2) in-city social interaction control and 3) intercity travel restrictions. These measures can be directly mapped to the three key parameters $\kappa_I$, $\kappa_E$, and $\Delta N$ (denoted as TL henceforth) in the STEx-SEIR model (Table S1), and thus empower the quantitatively evaluation of the effectiveness and cost-effectiveness of various scenarios.

### 2.2 Epidemiological Control Measures

Regular epidemiological control measures include *thorough* epidemiological surveys, close contact tracing, and quarantines of all the infected individuals and their close contacts as early and as possible—especially the close contacts, who may be potential carriers. These measures are reflected in the model by tuning the values of $\kappa_I$ and $\kappa_E$.

It is noteworthy that the limits of the effectiveness of regular epidemiological control measures:

- It is very difficult for the close-contact tracing and quarantines to be *exhaustive*, so they cannot reduce $\kappa_E$ to 0, indicating an effectiveness "roof";
- Close contact tracing is a tedious work and is heavily manpower dependent. Given that the epidemiological control system in regular practice usually has limited capacity. Under outbreaks the system could be overloaded and becomes dysfunctional. Under these circumstances, more aggressive measures are needed to reduce $\kappa_E$.

### 2.3 In-city Social Interaction Control Measures

In-city social interaction control measures include work-from-home, shutdown of schools, workplaces, public transit, and public spaces, cancelation of public events, lockdown of residential neighborhood, etc. The effects of these measures are to reduce the chance of close contact between the infected/exposed and susceptible individuals, and are represented in the model by the values of $\kappa_I$ and $\kappa_E$, especially $\kappa_E$, with contributions of different measures *aggregated* to a final value. We collected the measures adopted in different prior papers and coded them to estimate the final values of $\kappa_I$ and $\kappa_E$. It needs to be noted that the effects of specific measures are clearly *overlapping* with each other. For example, when social distancing both at the work and home ends are effectively implemented, the marginal "contribution" of an additional control on public transit usage would be minimal. Also, despite of the *nominal* public orders, the actual stringency of the measures could be *flexible* in practice. For simplicity, we utilize expert knowledge to estimate the overall efforts, i.e., the values of $\kappa_I$ and $\kappa_E$.

### 2.4 Intercity Travel Restrictions

Intercity travel restriction includes suspending inter-provincial and intercity buses, trains and flights, closing highways and roads, etc. The purpose is to reduce the intensity of personnel exchanges across cities, thereby reducing the transmission of the infected/exposed individuals. This is represented in the model by the reduction of spatial interaction parameter TL, or $\Delta_E$ and $\Delta_I$.

### 2.5 Evaluating Effectiveness of Antiepidemic Measures

In this paper we use two indicators to evaluate the effectiveness of disease control measures: the number of infected cases, and the spatial extent (represented by the number of cities with infected cases).

#### 2.5.1 Number of Infected Cases

Minimizing the total infected population is a self-explanatory goal from the perspective of epidemic control. Since *E* will eventually be converted to *I* with a fixed probability, we only need to focus on the infected population *I* and use the relative change of *I* over the simulation period (one month) as one indicator of effectiveness.

$$\mathrm{Effect}_{\mathrm{I}_i} = \frac{I_i(t+30) - I_i(t)}{I_i(t)} \times 100\%$$

The total antiepidemic effect in all cities is the summation:

$$\mathrm{Effect}_I = \sum\nolimits_i \mathrm{Effect}_{\mathrm{I}_i}$$

#### 2.5.2 Number of Cities with Infected Cases

Aside from the total number of infected cases, the spatial extent of the epidemic is also an indicator worth noticing. If the epidemic can be contained within a narrow spatial extent, the medical resources in other cities can be *diverted* to the infected cities to help, which actually happened in China where medical teams across the country were sent to Wuhan for support. We also use the relative change of the number of cities with infected cases during the simulation period (one month) to quantify the effect:

$$\mathrm{Effect}_{\mathrm{C}_i} = \frac{C_i(t+30) - C_i(t)}{C_i(t)} \times 100\%$$

The total effect in all cities is the summation:

$$\mathrm{Effect}_C = \sum\nolimits_i \mathrm{Effect}_{\mathrm{C}_i}$$

### 2.6 Cost of Antiepidemic Measures

We used the *opportunity cost of economic output* as the indicator for the cost of the antiepidemic measures. For simplicity, we only used the work-hour loss as a proxy for the economic output loss and did not convert it to monetary values of goods and services.

#### 2.6.1 Cost of Travel Restriction

Travel restriction leads to reduction in the total number of personnel exchanges between cities. In particular, the outbreak of COVID-19 in China overlapped with the Spring Festival holidays, and travel restriction apparently caused the migrant workers who previously travelled back hometown not be able to return to cities where they had worked after the holidays. While the impact on economic output is affected by factors such as the productivity difference between the migrant workers and local workers, we can use wages as a proxy for productivity and estimate the differences (*32–34*).

Calculations show that the average income of migrant workers in 2017 was 100.36% of that of the local workers, or roughly the same, so we do not need to adjust for the differences in the model[1]. Therefore, this study ignores the role of the above factors. Factors such as employment elasticity, which are more difficult to estimate, are also omitted for simplicity. Therefore, the loss of economic output of city *i* at time *t* from travel restriction is:

$$\text{Inter_Cost}_i(t) = (N\text{_normal}_i(t) - N_i(t)) \times \kappa_E\text{_}n\text{ormal}_i(t)$$

where $N$_normal is the urban population in the normal status, approximated by the number of urban populations in the same period last year, and $\kappa_E$_normal is the $\kappa_E$ in the normal status.

### 2.6.2 The Cost of In-city Social Interaction Control Measures

For those who are not affected by the intercity travel restriction, we argue that the indiscriminately implemented in-city social interaction control measures reduce their intensity of social interaction, which in turn reduces their productivity and thus reduces the economic output. More specifically, we used $\kappa_E$ and $\kappa_I$ as the proxy for productivity.

It should be noted that lower $\kappa_E$ or $\kappa_I$ is not necessarily the result of indiscriminate suppression of daily interactions but could also be the result of regular epidemiological control measures. A distinction needs to be made between these two situations. It can be reasonably assumed that when the number of the infected cases is small. For example, the daily increase of the number of infected cases does not exceed a certain threshold, the goal of reducing $\kappa_I$ and $\kappa_E$ can be achieved *solely* by regular epidemiological control measures. Therefore, a rational administration will choose not to opt to universal in-city social interaction control measures to avoid collateral loss. What is lost in this case is the output of only those who have been quarantined, i.e., the exposed and infected cases.

We admit that it is extremely difficult to estimate such a threshold, which could be highly dependent on specific contexts such as governance capability, the resourcefulness of the epidemic control system, and cultural and geographic factors. Although, a numerical estimation would still be of help. There was no opportunity to observe this threshold during the outbreak in China: in the early stage of the epidemic (before January 23, 2020), no effective epidemic control practices were taken outside of Wuhan, and after January 23, the country had generally turned to aggressive and universal antiepidemic measures and it is difficult to separate the net effect of the each epidemic control measure. However, the Singapore situation in January and February may constitute a good example, where in most cases daily reported new infected cases were under 10. Another remark is that the Chinese government sets a threshold of 3 cases for reporting a newly discovered epidemic to the national CDC. In light of these numerical example, we use 10 daily new infected cases as an estimate of the capacity threshold value. Therefore, the loss of economic output of city *i* at time *t* from in-city social interaction control measures is set as follows.

$$\text{Intra_Cost}_\text{i}(t) = \begin{cases} S_i(t) \times \left(\kappa_E\text{_normal}_i(t) - \kappa_{E_i}(t)\right), & \alpha E_i(t) > 9 \\ (N_i(t) - S_i(t)) \times \ (\kappa_E\text{_normal}_i(t) - \kappa_{E_i}(t)), & \alpha E_i(t) \leq 9 \end{cases}$$

[1] The monthly wage level of the migrant population is from the result of the "2017 China Migrant Population Dynamics Monitoring Survey." There are 170,000 samples, which are nationally representative. The calculation process uses weighted PPS sampling. The average urban wage is calculated by weighting the employment based on the urban private and non-private wage level.

### 2.6.3 Comprehensive Cost Function

For straightforwardness, we convert the economic output loss into to a relative value: the proportion of the output with regard to that of the baseline (the normal situation). The total output loss ratio of city $i$ during the entire simulation period is

$$\text{Cost}_i = \frac{\sum_{t=0}^{n}(\text{Inter_Cost}_i(t) + \text{Intra_Cost}_i(t))}{\sum_{t=0}^{n}(N_\text{normal}_i(t) \times \kappa_E_\text{normal}_i(t))} \times 100\%$$

and the ratio of total output loss of all cities during the entire simulation period is:

$$\text{Cost} = \frac{\sum_i \sum_{t=0}^{n}(\text{Inter_Cost}_i(t) + \text{Intra_Cost}_i(t))}{\sum_i \sum_{t=0}^{n}(N_\text{normal}_i(t) \times \kappa_E_\text{normal}_i(t))} \times 100\%$$

## 2.7 Comprehensive Cost-effectiveness of Antiepidemic Measures

In summary, the comprehensive cost-effectiveness function of antiepidemic measures is:

$$\text{Cost/Effectiveness}_I = \frac{\text{Effect}_I}{\text{Cost}}$$
$$\text{Cost/Effectiveness}_C = \frac{\text{Effect}_C}{\text{Cost}}$$

Particularly, if Cost = 0, it is stipulated that both $\text{Cost/Effectiveness}_I$, $\text{Cost/Effectiveness}_C$ equal to 0.

## 2.8 Scenario Design for Epidemic Control Measures

For simplicity, we only calculated the cost and cost-effectiveness functions at the national level (the entire city network). To discover any initial value-dependence of the cost and the cost-effectiveness functions, we ran the simulation at different stages of the epidemic. Based on the calibrated model, five scenarios with the following starting and ending dates are set:

- Early Stage: December 31, 2019–January 29, 2020
- Accelerating Stage: January 16, 2020–February 15, 2020
- Peak/Inflecting Stage: February 5, 2020–March 5, 2020
- Declining Stage: February 28, 2020–March 28, 2020
- Ending Stage: March 16, 2020–April 14, 2020

The model is run for 30 days for each scenario, and five indicators are calculated: number of total infected cases, number of cities with infected cases, overall cost, effectiveness/cost ratio with respect to the number of infected cases, and effectiveness/cost ratio with respect to the number of cities with infected cases.

We calculated the gradient function of the five indicators with respect to $\kappa_E$, $\kappa_I$, and TL to analyze the structure of the solution space. For simplicity, we calculate the gradients discretely at the following points: $\kappa_I$ = {0, 1, 2, 3, 4, 5, 6, 7, 8, 9}; $\kappa_E$ = {0, 2, 4, 6, 8, 10, 12, 14, 16}; TL = {1, 2, 3, 4, 5}, where the five levels of TL correspond to 0.2, 0.4, 0.6, 0.8, and 1 times of the *normal* intercity travel level. Note that the lowest level of TL is not 0. This is because according to the situation in China (given by Baidu Migration Data), when the TL is the lowest, the travel level is still about 20% of the normal level, and we believe that the stringency of intercity travel restriction in reality cannot be greater than this level. In contrast, aggressive in-city social interaction control measures can reduce the values of $\kappa_E$ and $\kappa_I$ to levels very close to 0.

## 3. Antiepidemic Situation Assessment of Countries/Regions

We evaluated the antiepidemic situation of countries/regions based on the stringency of activity control policies of the infected and exposed, i.e., the degree of which $\kappa_I$ and $\kappa_E$ are reduced. This takes 3 steps: 1) codifying policies and computation of the stringency indices; 2) mapping of the stringency indices into values of $\kappa_I$ and $\kappa_E$; and 3) assessment of the antiepidemic situation based on the results of our analysis.

### 3.1 Codifying Policies and Computing Stringency Indices

For the exposed, effective activity control measures include close-contact tracing and social distancing. For the computation of the stringency of these measures, we used the data and method provided by the Oxford COVID-19 government response tracker (*17*), and computed and compiled the exposed activity control stringency index with six indicators: school shutdown, workplace shutdown, public events cancelation, public transport shutdown, public information campaign, and close-contact tracing.

For the infected population, quarantine is the most important measure to reduce their activities, which but also requires testing in the first place. We therefore used the *comprehensiveness of testing* as the proxy for the infected activity control stringency. As testing policy coding from the aforementioned source is rather coarse, we instead designed an alternative index based on two indicators: tests per 1 million population, and the ratio between cumulative cases and the number of total tests. The compiled index is gained through dividing the first indicator with the second, logarithmically transforming the quotient, and finally normalizing the result to the 0-100 range to match that of the exposed population's activity control stringency index. The logarithmic transformation is used only for visualization purposes, as very few countries/regions have much higher original indices. For the testing data, we employed sources from www.worldometers.info/coronavirus/. It should be noted that as mainland China does not publish testing data, we used the data from Wuhan as a proxy.

For both indices, we use the data as of April 7. A total of 99 countries/regions are present in the final dataset.

### 3.2 Mapping of the stringency indices into $\kappa$ values, and division criteria of the three strategy groups

Analysis results show that there exist two threshold values for both $\kappa_I$ and $\kappa_E$ that correspond to the "elimination" ($R_0 << 1$) and "control" ($R_0 < 1$) strategies, as shown in Table S2. It should be noted that the effects of the two indicators are substitutable, with the elasticity of $\kappa_I$ to be larger.

For the mapping from $\kappa_I$ to the respective stringency index of a specific measure, it is clear that the mapping should not be linear because of the non-linear transformations we have done. We therefore employed an estimation method based on expert knowledge: to achieve 2/3 reduction of $\kappa_I$ which is required to "eliminate" the epidemic, it is needed to quarantine as many infected cases as possible. Considering that about 1/3 to 1/2 infected cases are asymptomatic (*35*), this roughly means that all symptomatic cases should be tested. Based on the statistics from Wuhan and South Korea, this requires the tests per 1 million population to be greater than 10,000, and the ratio between the number of cumulative cases and total tests to be smaller than 5%.

Converted from the above estimation to the respective stringency index, the value is around 70. We thus used this value as the boundary between "elimination" and "control". Following similar methods, we estimated that the boundary between the "control" and "delay" policy groups to be around 60.

For the mapping from $\kappa_E$ to the respective stringency index, because all six indicators' contribution to the actual stringency level are roughly equally weighted, we simply conducted a linear mapping between the composite stringency index and $\kappa_E$. Thus, the stringency index boundary between the "elimination" and "control" groups is 75, and that between "control" and "delay" is 50.

## 4. Data Processing and Parameter Estimation

### 4.1 Estimation of Epidemiological Parameters

Epidemic parameter estimation is the first and fundamental step in the SEIR model. In previous studies, choices of these parameters usually adopted mathematical interpretation, such as the inverse of the incubation period or infectious period (*7*, *23*) and thus the numbers are not physically interpretable. However, if we consider the actual characteristics of COVID-19 as shown in epidemiological studies, the ambiguity of the model can be greatly reduced. We estimated these parameters based on real-world data, considering errors caused by possible concealment, omission, and late reports.

#### 4.1.1 Transmission Rates from Susceptible to Infected ($\boldsymbol{\beta_I}$) and Exposed ($\boldsymbol{\beta_E}$)

We used close contact tracing data after January 28 to estimate $\beta_I$, since statistical data in the early stage showed fluctuation due to the lag of reactions from the government. In the equation following, $I(t)$ indicates the number of newly confirmed cases in a day. And $C(t)$ indicates the number of close contacts who are still under medical observation in the same day. We averaged $\beta_I(t)$ with regard to *t* and got 0.01986, so we set $\beta_I$ to 2%.

$$\beta_I(t) = \frac{I(t)}{C(t)}$$

In addition, we considered COVID-19 to be infectious in the later stage of the incubation period: there is no evidence of weaker infectivity during the incubation period, and it is assumed that it can be as infectious as during the infected period, indicating $\beta_E = \beta_I = 2\%$. The incubation period is generally 2-14 days, and 3-7 days in most cases. The median incubation period is 4 days. Assuming infectivity in the 1-3 days before the end of the incubation period, with an average of 2 days, the ratio of infectious exposed people ($\theta_E$) can be calculated as follows

$$\theta_E = \frac{\overline{E_\iota}}{\overline{E}}$$

where $\overline{E_\iota}$ is the average infectious day of the exposed, and $\bar{E}$ is the median of the incubation period.

### 4.1.2 Ratio of Exposed Cases Turning to Infected Cases: $\alpha$

According to the *New Coronavirus Pneumonia Prevention and Control Program* (Second Edition), suspected cases are those with similar clinical symptoms as infected people. If a suspected case tests positive for nucleic acid test or its viral gene sequencing is highly homologous with SARS-CoV-2, it is diagnosed as a confirmed case.

Considering the relationship between suspected and confirmed cases, the ratio between new confirmed cases to existing suspected cases is recorded as the ratio of the exposed cases becoming infected cases. This index is a characteristic of the disease itself and is generally the inverse of the incubation period from the mathematical interpretation. Therefore, it should be uniform for all cities. This parameter has converged since January 28 and the average value is 0.1404 (14.04%).

Because of the presence of asymptomatic infections, only considering suspected cases can lead to overestimation. Since asymptomatic infections accounted for 1.2% of confirmed cases, we scaled up the suspected cases and the revised $\alpha$ to be 13.87%.

### 4.1.3 Mortality Rate

According to the Epidemiology Working Group of the Chinese Center for Disease Control, at least 104 cases existed in Hubei Province in December 2019, spread in fourteen counties. By January 10, 2020, there were 757 cases and 102 deaths in 113 cities of 20 provinces. However, in the national epidemic reporting system, the first death was not reported until January 20 in the whole country except Wuhan. Therefore, we took the results (Table S3) published by the Epidemiology Working Group and calculated the mortality rates in different regions at each stage (Table S4).

We fitted the mortality rates with the following functions:

Wuhan:

$$\nu = 2.337e^{-1.117t} - 4.942e^{-2.075t}$$

Hubei province except Wuhan:

$$\nu = (2.337e^{-1.117t} - 4.942e^{-2.075t}) * 0.69$$

Mainland China except Hubei province:

$$\begin{cases} \nu = 0\,, t < t_{\text{first}} \\ \nu = 0.05046t^{-2.368}\,, t \geq t_{\text{first}} \end{cases}$$

where $t_{\text{first}}$ indicates January 1, 2010, which means that there were no deaths in any region of China except Hubei province in December 2019. The above fitting all passed the goodness-of-fit tests.

### 4.1.4 Recovery Rate

Recovery rate is closely related to the capacity of local medical resources and city governments' responsiveness, so we conducted parameter fitting for each city. It is reasonable to believe that the number of cured cases published is reliable since there is no obvious reason for concealing recovery.

Therefore, the recovery rate is calculated as the ratio between newly cured cases and existed confirmed cases. The recovery rate has changed significantly by time. Taking Wuhan as an example, the abrupt outbreak led to a lack of medical resources at the beginning. Then, aid resources from all over the world were sent to Hubei province. In addition, *Huǒshénshān* and *Léishénshān* Hospitals were built quickly to treat patients in severe conditions. These efforts contributed to the improvement of recovery rates (*7*). To capture the time-wise changes, we fitted piecewise functions to the recovery rates based on the characteristics of each stage.

Among the different stages, the early recovery rate in Wuhan fluctuated too heavily to fit, so we took the average value in the period as an indicator of the recovery rate.

The specific fitted functions are as follows:

Mainland China except Hubei province:

$$\begin{cases} \mu = 0\,, t < t_{\text{control}} - 1 \\ \mu = 0.1215e^{-1.572t} + 0.0033e^{0.1188t}\,, t_{\text{control}} - 1 \le t < t_{\text{stage}} \\ \mu = 0.04202e^{0.01073t} + 0.00009881e^{1.08t}, t_{\text{stage}} \le t < t_{\text{stage}} + 5 \\ \mu = 0.00112t^{1.58205} + 0.080136\,, t \ge t_{\text{stage}} + 5 \end{cases}$$

Hubei province except Wuhan:

$$\begin{cases} \mu = 0\,, t < t_{\text{control}} + 2 \\ \mu = 0.03674e^{-1.359t} + 0.00015e^{0.1433t}\,, t_{\text{control}} + 2 \le t < t_{\text{predict_start}} \\ \mu = 0.012286t^{0.71869} + 0.0794335, t \ge t_{\text{predict_start}} \end{cases}$$

Wuhan:

$$\begin{cases} \mu = 0.0576 \quad, t < t_{\text{control}} \\ \mu = 0.013696 \quad, t_{control} \le t < t_{\text{stage}} \\ \mu = 0.03271e^{0.08541t} - 3.3562e^{-5.449t}, t_{\text{stage}} \le t \end{cases}$$

where $t_{\text{contro}l}$ is the time of Wuhan lockdown: January 23, 2020; $t_{\text{stage}}$ is February 15, 2020; and $t_{\text{predict_start}}$ is the last day of collecting reported data for this paper: February 26, 2020.

### 4.2 Estimating Parameters of Social Interaction Intensity

Close contact tracing is a common epidemiological control measure, especially when early treatment is not clear, and when vaccines are not available. This approach can identify potentially infected individuals, quickly isolate them before they turn to severe patients, and prevent the occurrence of secondary transmission (*36*).

Social interaction intensities of the exposed and infected individuals are denoted as $\kappa_E$ and $\kappa_I$ respectively in our model. These two parameters are highly related to local population densities. Considering that the COVID-19 outbreak occurred during the Chinese New Year, the intensity of social interaction was significantly higher than usual.

Wuhan, as China's large transportation hub, may have a higher per capita social interaction intensity than other cities. To address this, we used the contact rate of the infected population during SARS (maximumly 12) as the initial value of $\kappa_I$ (*37*).

In general, the behavior of the exposed population is less restricted than the infected, so we assume that the maximum contact rate of the exposed is twice as high as that of the infected, which means that the maximum $\kappa_E$ is 24.

According to the local epidemic reports, the initial values of $\kappa_I$ and $\kappa_E$ in each area can be obtained (Table S5). Then, based on the local control policies and response time, we adjusted the contact values at different stages. In January 20, 2020, Zhong Nanshan announced that there was a risk of human-to-human transmission of COVID-19, and some people in Wuhan began to consciously reduce their activities. Until the citywide lockdown, the number of cases surged, and people began to realize the severity of the situation. Then, strategies such as business shutdown, school shutdown, stay-at-home notices, and community shutdown were gradually implemented throughout the country to avoid contact between the susceptible and the infected population.

In the late stage of the epidemic, numerous prevention and control strategies, including hospitalizing all confirmed cases and isolating all suspected cases, have greatly reduced the values of $\kappa_I$ and $\kappa_E$ (Figs.1 and 2).

### 4.3 Urban Population and Human Interaction Data

The initial value of each city's population in our model is that on December 8. Then, the population changed after the New Year migration, which can be calculated through the spatial interaction matrix.

Our research includes 368 prefectures in mainland China, with the time span from December 1, 2019 to April 30, 2020. The data collected included migration data from the Baidu Migration Map (*15*), statistical yearbooks, and local COVID-19 reports. The original Baidu migration data is consisted of migration flows of 91,788 city pairs among 368 cities from January 1, 2020 to February 26, 2020.

Since there was no news of the COVID-19 outbreak in December, the intensity of social activities was not affected and remained the same as normal workdays. In March and April after the outbreak, we assumed that due to the impact of the travel restrictions in most cities, residents' travel remained at a very low level.

Therefore, the migration data for December was set to be the same as an average workweek from January 6, 2020 to January 12, 2020, and the intercity migration after February 26 was set to be the same as between February 17 to February 23. In addition, the numbers of COVID-19 cases in our model and population data for the 368 cities were collected from the National Bureau of Statistics of China (*38*) and the National Health Commission of China's reports.

### 4.4 Codifying Local Epidemic Control Policies

We collected the publicly announced epidemic control policies of all prefectures in China from January 23 to February 20. This dataset includes about 4,000 documents, among which 364 were issued at the provincial level. We codified the policies into 6 categories: one for long-distance travel restrictions, and the other five for in-city social interaction control measures (Table S6). We further translated the policies into their control stringency in terms of social interaction reduction based on expert evaluation on the policy terms as well as enforcement level inferred from media coverage. We lastly mapped the codified policies to the values of $\kappa_I$ and $\kappa_E$ (Figs. 1 and 2).

## 5. Cost-Effectiveness Analysis Specification

### 5.1 Early Stage: December 31, 2019 to January 29, 2020

#### 5.1.1 Effectiveness of Antiepidemic Measures

In the baseline (real-world) model, the initial number of cases at this stage was 53 and increased to 7,780 after 30 days. The initial number of infected cities were 23 and increased to 301 after 30 days.

In terms of the reduction of infected cases, the decreases of the three coefficients, $\kappa_I$, $\kappa_E$, and TL all have positive effects, but the effects are non-linear—with all returns diminishing marginally. The marginal contribution of $\kappa_I$ and $\kappa_E$ are roughly comparable and also substitutable, with that of $\kappa_E$ slightly higher when the level of social interactions is high (i.e., the values of $\kappa_I$ and $\kappa_E$ are high, or close to the ordinary level, which is 9 for $\kappa_I$ and 16 for $\kappa_E$), and that of $\kappa_I$ higher when the level of social interactions is low (as low as 0). However, the marginal contribution of $\kappa_I$ and $\kappa_E$ are both 2-5 times higher than that of TL in most cases, and the latter is only significant when the values of $\kappa_I$ and $\kappa_E$ are both very low. When $\kappa_I \le 3$ and $\kappa_E \le 4$, the number of infected cases will shrink, and will be reduced to 0 after approximately two months (Table S7).

In terms of the reduction of number of cities with infected cases, the decreases of the three coefficients, $\kappa_I$, $\kappa_E$, and TL all have positive and non-linear effects, but with increasing returns marginally. The marginal contribution of $\kappa_I$ and $\kappa_E$ are roughly comparable and also substitutable, with that of $\kappa_E$ slightly higher when the level of social interactions is high, and that of $\kappa_I$ slightly higher when the level of social interactions is low. However, the marginal contribution of $\kappa_I$ and $\kappa_E$ are both 1–2 times higher than that of TL in most cases, and the latter is only significant when the values of $\kappa_I$ and $\kappa_E$ are both very low. When $\kappa_I \le 3$ and $\kappa_E \le 4$, the number of cities with infected cases will shrink, and will be reduced to 0 after approximately two months.

#### 5.1.2 Cost-Effectiveness of Antiepidemic Measures

Since there was no baseline control at this stage, any change in the three coefficients means tighter control on social interactions, and hence a cost to economic output.

First, the marginal contribution from the reduction of TL to the cost function is approximately 2–5 orders of magnitude lower than that of $\kappa_I$ and $\kappa_E$ such that its impact is mostly negligible.

Second, the cost function has a ridge along the direction of ($\kappa_I = 9$, $\kappa_E = 12$) and ($\kappa_I = 0$, $\kappa_E = 16$). The global gradient between the ridge and ($\kappa_I = 0$, $\kappa_E = 0$) is moderate, and a lowest plateau appears when $\kappa_I \le 6$ where the value of the cost function is 0. On the other side, the global gradient is much steeper, with a minimal value of the cost function that is also 0 at the corner. The global cost peak appears at ($\kappa_I = 9$, $\kappa_E = 12$), at which point the peak cost slightly exceeds 4% of the total output.

In terms of cost-effectiveness, as the peaks and trends of the cost function and two effect functions are different, the two cost-effectiveness functions show greater non-linearity. In terms of the cost-effectiveness with respect to the reduction of the number of infected cases, two local peaks exist. One is at the "strictest control" point i.e., ($\kappa_I = 0$, $\kappa_E = 0$), when each 0.3% reduction in the number of infected cases is associated with 1% of economic output cost within a month; another peak occurs at the "non-control" point i.e., ($\kappa_I = 9$, $\kappa_E = 16$), when each 1.7% reduction in the number of infected cases is associated with 1% of economic output cost within a month.

However, in the latter case, there will be *no* reduction in the number of infected cases at all according to the respective effectiveness function, while the former case coincides with the effectiveness peak. In terms of the cost-effectiveness with respect to the reduction of the number of cities with infected cases, situations are similar, only that a third peak appears at ($\kappa_I = 0$, $\kappa_E = 16$), which is also a hardly effective solution. Overall, the best solution considering both effectiveness and cost-effectiveness in whichever terms at this stage of epidemic is at the "strictest control" point (Fig. S3).

## 5.2 Accelerating Stage: January 16, 2020 to February 15, 2020

### 5.2.1 Effectiveness of Antiepidemic Measures

In the baseline (real-world) model, the initial number of cases at this stage was 1,059 and increased to 8,920 after 30 days. The initial number of infected cities were 178 and increased to 310 after 30 days.

In terms of the reduction of infected cases, the decreases of the three coefficients, $\kappa_I$, $\kappa_E$, and TL all have positive effects, but the effects are non-linear—with all returns diminishing marginally. The marginal contribution of $\kappa_I$ and $\kappa_E$ are roughly comparable and also substitutable, with that of $\kappa_E$ slightly higher when the level of social interactions is high (i.e., the values of $\kappa_I$ and $\kappa_E$ are high, or close to the ordinary level, which is 9 for $\kappa_I$ and 16 for $\kappa_E$), and that of $\kappa_I$ higher when the level of social interactions is low (as low as 0). However, the marginal contribution of $\kappa_I$ and $\kappa_E$ are both up to two orders of magnitude higher than that of TL in most cases, and the latter is only significant when the values of $\kappa_I$ and $\kappa_E$ are both very low. When $\kappa_I \leq 3$ and $\kappa_E \leq 4$, the number of infected cases will shrink, and will be reduced to 0 after approximately two months (Table S8).

In terms of the reduction of number of cities with infected cases, the decreases of the three coefficients, $\kappa_I$, $\kappa_E$, and TL all have positive and non-linear effects, but with increasing returns marginally. The patterns of marginal contribution of the three coefficients are similar with the "reduction of infected cases" case and we do not elaborate here.

### 5.2.2 Cost-Effectiveness of Antiepidemic Measures

First, the marginal contribution from the reduction of TL to the cost function is approximately 2–4 orders of magnitude lower than that of $\kappa_I$ and $\kappa_E$ such that its impact is mostly negligible.

Second, the cost function has a ridge along the direction of ($\kappa_I = 9$, $\kappa_E = 4$) and ($\kappa_I = 0$, $\kappa_E = 16$). The global gradient between the ridge and ($\kappa_I = 0$, $\kappa_E = 0$) is moderate and roughly linear, and a lowest plateau appears when $\kappa_I \leq 3$ and $\kappa_E \leq 4$ (the "strictest control" scenario) where the cost is a small number (2%–4%). On the other side, the global gradient is much steeper with increasing margins, with a minimal value of the cost close to 0 at the corner (the "non-control scenario). One noteworthy character of the cost function is that the value of $\kappa_I$ does not affect the position of the peak, but affects its value: keeping $\kappa_E$ the same, each 10% reduction in $\kappa_I$ reduces the cost function by 2%–4%. The global cost peak appears at ($\kappa_I = 9$, $\kappa_E = 4$), or a "loose control" scenario, at which point the peak cost exceeds 40% of the total output. The intuition is that in this case, the control on social interactions is not sufficient to reverse the trend curve of the epidemic, and after the number of infected cases exceeding a certain threshold (the capacity "roof" of regular epidemiological measures to contain the epidemic), cities have to opt to a universal control on everyday social interactions to contain the epidemic, inflicting heavy loss on economic output.

The global gradients of the cost-effectiveness function for both effectiveness metrics show similar patterns with the previous stage, only with more non-linearity, which is too complicated to be elaborated here. For more details, refer to Fig. S4.

## 5.3 Peak Stage: February 5, 2020 to March 5, 2020

### 5.3.1 Effectiveness of Antiepidemic Measures

The baseline model witnessed the turning point of the epidemic, when the initial number of cases at this stage was 10,345 and reduced to 1,913 after 30 days. The initial number of infected cities were 314 and reduced to 253 after 30 days.

In terms of the reduction of infected cases, the decreases of $\kappa_I$ and $\kappa_E$ both have positive effects, but the effects are non-linear—with all returns diminishing marginally. The decrease of TL, however, have varying directions of effect dependent on the values of $\kappa_I$ and $\kappa_E$, with negative effect when both $\kappa_I$ and $\kappa_E$ values are high and positive otherwise. The marginal contribution of $\kappa_I$ and $\kappa_E$ are roughly comparable and also substitutable, with that of $\kappa_E$ slightly higher when the level of social interactions is high (i.e., the values of $\kappa_I$ and $\kappa_E$ are high, or close to the ordinary level, which is 9 for $\kappa_I$ and 16 for $\kappa_E$), and that of $\kappa_I$ one order of magnitude higher when the level of social interactions is low (as low as 0). However, the marginal contribution of $\kappa_I$ and $\kappa_E$ are both up to two orders of magnitude higher than that of TL in most cases, and the latter is only significant when the values of $\kappa_I$ and $\kappa_E$ are both very low. When $\kappa_I \leq 3$ and $\kappa_E \leq 6$, the number of infected cases will shrink, and will be reduced to 0 after approximately two months. Higher values of $\kappa_I$ and $\kappa_E$ (looser control on the level of social interactions) will not be sufficient to reverse the epidemic trend curve. And at the non-control scenario, the number of infected cases will grow to approximately 200,000, or 20 times the number at the beginning of the baseline scenario (Table S9).

In terms of the reduction of number of cities with infected cases, the decreases of the three coefficients, $\kappa_I$, $\kappa_E$, and TL all have positive and non-linear effects, but with increasing returns marginally. The patterns of marginal contribution of the three coefficients are similar with the "reduction of infected cases" case and we do not elaborate here.

### 5.3.2 Cost-Effectiveness of Antiepidemic Measures

First, the marginal contribution from the reduction of TL to the cost function is approximately 2–4 orders of magnitude lower than that of $\kappa_I$ and $\kappa_E$ such that its impact is mostly negligible.

Second, the cost function has a ridge along the direction of ($\kappa_I = 9$, $\kappa_E = 0$) and ($\kappa_I = 2$, $\kappa_E = 16$). The global gradient between the ridge and ($\kappa_I = 0$, $\kappa_E = 0$) is moderate, and the lowest cost at the corner (the "strictest control" scenario) is a small number (2%–4%). On the other side, the global gradient is much steeper, with a minimal value of the cost function close to 0 at the corner (the "non-control scenario). Again, the value of $\kappa_I$ does not affect the position of the peak but affects its value: keeping $\kappa_E$ the same, each 10% reduction in $\kappa_I$ reduces the cost function by 8%–9%. The global cost peak appears at ($\kappa_I = 9$, $\kappa_E = 0$), or a "loose control" scenario, at which point the peak cost exceeds 90% of the total output.

The global gradients of the cost-effectiveness function for both effectiveness metrics show similar patterns with the previous stage, only with more non-linearity, which is too complicated to be elaborated here. For more details, refer to Fig. S5.

## 5.4 Declining Stage: February 28, 2020 to March 28, 2020

### 5.4.1 Effectiveness of Antiepidemic Measures

In the baseline model, the initial number of cases at this stage was 3,886 and reduced to 22 after 30 days. The initial number of infected cities were 288 and reduced to 1 (Wuhan) after 30 days.

In terms of the reduction of infected cases, the decreases of $\kappa_I$ and $\kappa_E$ both have positive effects, but the effects are non-linear—with all returns diminishing marginally. The decrease of TL, however, have varying directions of effect dependent on the values of $\kappa_I$ and $\kappa_E$, with negative effect when both $\kappa_I$ and $\kappa_E$ values are high and positive otherwise. The marginal contribution of $\kappa_I$ and $\kappa_E$ are roughly comparable and also substitutable, with that of $\kappa_E$ slightly higher when the level of social interactions is high (i.e., the values of $\kappa_I$ and $\kappa_E$ are high, or close to the ordinary level, which is 9 for $\kappa_I$ and 16 for $\kappa_E$), and that of $\kappa_I$ 4 times higher when the level of social interactions is low (as low as 0). However, the marginal contribution of $\kappa_I$ and $\kappa_E$ are both up to 1 order of magnitude higher than that of TL in most cases, and the latter is only significant when the values of $\kappa_I$ and $\kappa_E$ are both very low. When $\kappa_I \leq 9$ and $\kappa_E \leq 10$, or $\kappa_I \leq 6$ and $\kappa_E \leq 12$, the number of infected cases will shrink, and will be reduced to 0 after approximately two months. Higher values of $\kappa_I$ and $\kappa_E$ (looser control on the level of social interactions) will not be sufficient to reverse the epidemic trend curve. And at the non-control scenario, the number of infected cases will grow to approximately 10,000, or three times the number at the beginning of the baseline scenario (Table S10).

In terms of the reduction of number of cities with infected cases, the decreases of the three coefficients, $\kappa_I$, $\kappa_E$, and TL all have positive and non-linear effects, with increasing returns marginally at most times except when the values of $\kappa_I$ and $\kappa_E$ are both very low—a noteworthy character that is different with the symmetric accelerating stage. Another noteworthy character of the effectiveness function is the occurrence of plateaus when $\kappa_I \leq 9$, where the change of $\kappa_E$ and TL will not affect the spatial scope of the epidemic.

#### 5.4.2 Cost-Effectiveness of Antiepidemic Measures

First, the marginal contribution from the reduction of TL to the cost function is approximately 2–4 orders of magnitude lower than that of $\kappa_I$ and $\kappa_E$ such that its impact is mostly negligible.

Second, the cost function has a ridge along the direction of ($\kappa_I = 9$, $\kappa_E = 10$) and ($\kappa_I = 0$, $\kappa_E = 10$). The global gradient at both sides of the ridge drops close to 0 with increasing margins and plateaus when $\kappa_I$ is sufficiently low or high. Again, the value of $\kappa_I$ does not affect the position of the peak, but affects its value: keeping $\kappa_E$ the same, each 10% reduction in $\kappa_I$ reduces the cost function by 5%–8%, and when $\kappa_I < 6$, a plateau occurs where cost is minimal. The global cost peak appears at ($\kappa_I = 9$, $\kappa_E = 10$), or a "loose control" scenario, at which point the peak cost exceeds 20% of the total output.

The global gradients of the cost-effectiveness function for both effectiveness metrics show similar patterns with the previous stage, only with more non-linearity, which is too complicated to be elaborated here. For more details, refer to Fig. S6.

### 5.5 Ending Stage: March 16, 2020 to April 14, 2020

#### 5.5.1 Effectiveness of Antiepidemic Measures

In the baseline model, the initial number of cases at this stage was 267 and reduced to 1 after 30 days. The initial number of infected cities were 99 and reduced to 1 (Wuhan) after 30 days.

In terms of the reduction of infected cases, the decreases of the three coefficients, $\kappa_I$, $\kappa_E$, and TL all have positive effects, but the effects are non-linear—with all returns diminishing marginally. The marginal contribution of $\kappa_I$ and $\kappa_E$ are roughly comparable and also substitutable, with that of $\kappa_E$ about two times higher at all times. However, the marginal contribution of $\kappa_I$ and $\kappa_E$ are both up to one order of magnitude higher than that of TL in most cases, and the latter is only significant when the values of $\kappa_I$ and $\kappa_E$ are both very low. When $\kappa_E \leq 12$, or $\kappa_I \leq 6$ and $\kappa_E \leq 14$, the number of infected cases will shrink, and will be reduced to 0 after approximately two months. Higher values of $\kappa_I$ and $\kappa_E$ (looser control on the level of social interactions) will not be sufficient to reverse the epidemic trend curve (Table S11).

In terms of the reduction of number of cities with infected cases, the decreases of the three coefficients, $\kappa_I$, $\kappa_E$, and TL all have positive and non-linear effects, this time all with decreasing returns marginally – a noteworthy character that is different with the symmetric early stage. Plateaus also exist at this stage, when $(\kappa_I, \kappa_E)$ is to the left (smaller) side of (6, 0)–(0, 8), where any further change of the coefficients will not affect the spatial scope of the epidemic.

### 5.5.2 Cost-Effectiveness of Antiepidemic Measures

At this stage, the shape and value of the cost function are both dominated by $\kappa_E$. They decrease monotonically and roughly linearly along with the direction of $\kappa_E$'s decline, and the effects of $\kappa_I$ and TL are very small.

At this stage, the epidemic at the baseline has come to an end. Although the initial number of cases is roughly equivalent to the initial stage of the epidemic, the strict control at the previous stages has led to a significant reduction in the number of the exposed population. Therefore, the cost function at this time is dominated by $\kappa_E$, and the peak value of the cost function is only 0.5% of the output, a great reduction compared to those in the previous three stages.

Since the lift of controls on social interactions will help restore economic output, the peak cost-effectiveness at this stage all occurs when all controls are lifted, at which time economic output will be very close to the normal level (when there is no epidemic). The second highest cost-effectiveness peak appears when the controls are maintained (given the number of infected cases, the controls now are realized predominately through effective epidemiological measures, rather than universal containment of social interactions), where the maximum loss of economic output is about 0.5%. While the former appears tempting, in the long run (about two months later) it is likely to cause a rebound in the epidemic (we only run the simulation for one month, so this effect cannot be directly observed in the model. However, the situation is similar to the initial period of the epidemic, and thus the lift of controls is just equivalent to starting the entire epidemic from the beginning, such that a second outbreak is inevitable if other conditions remain unchanged). The latter, on contrary, appears to achieve a good balance between cost and effectiveness within this study's analytical framework (Fig. S7).

**Fig. S1. The $\kappa_I$ of 32 provinces and selected cities from December 8, 2019 to February 25, 2020.**

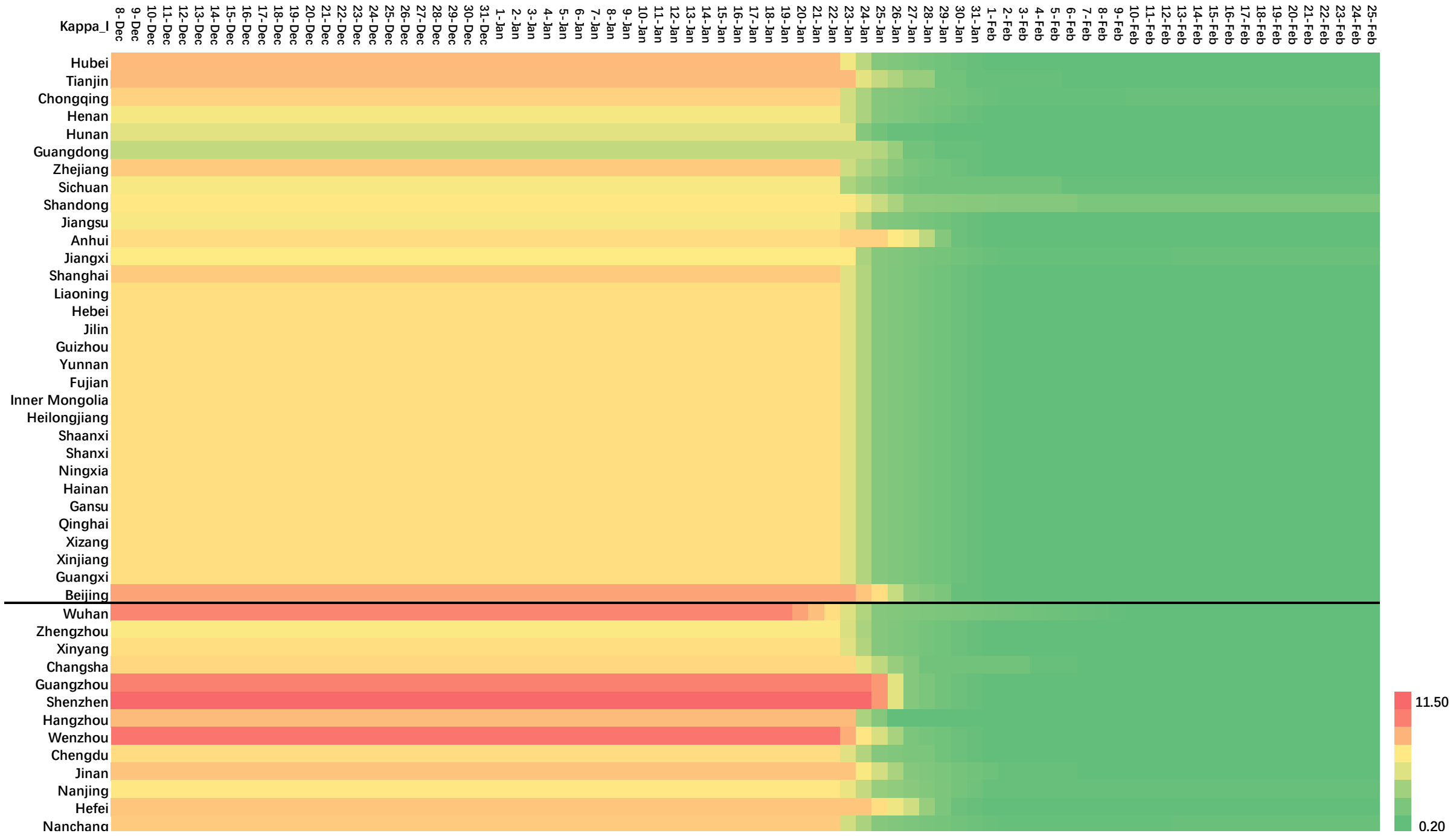

**Fig. S2. The $\kappa_E$ of 32 provinces and selected cities from December 8, 2019 to February 25, 2020.**

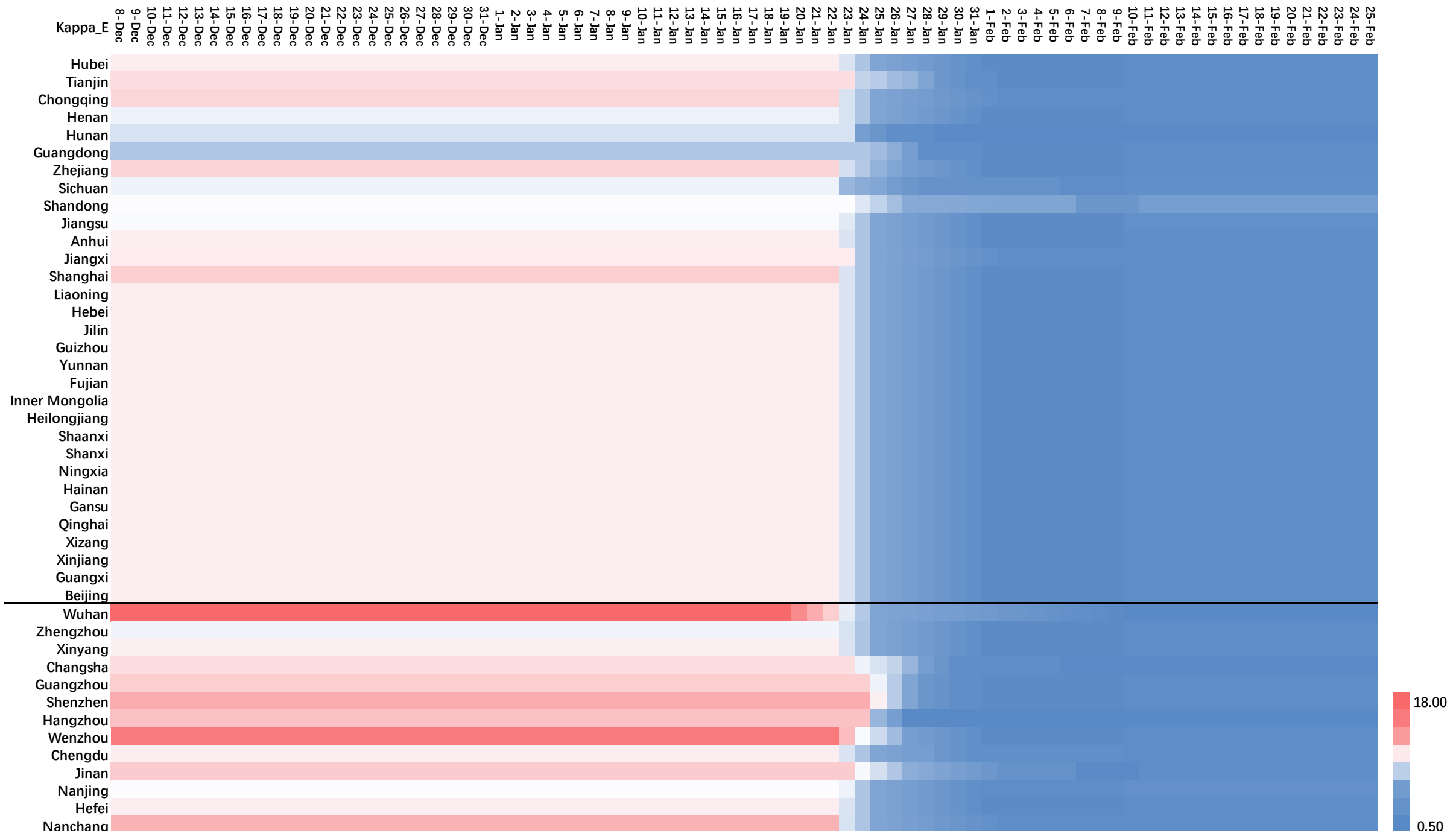

**Fig. S3. Global gradients of the cost, effectiveness, and cost-effectiveness functions at the early stage.**

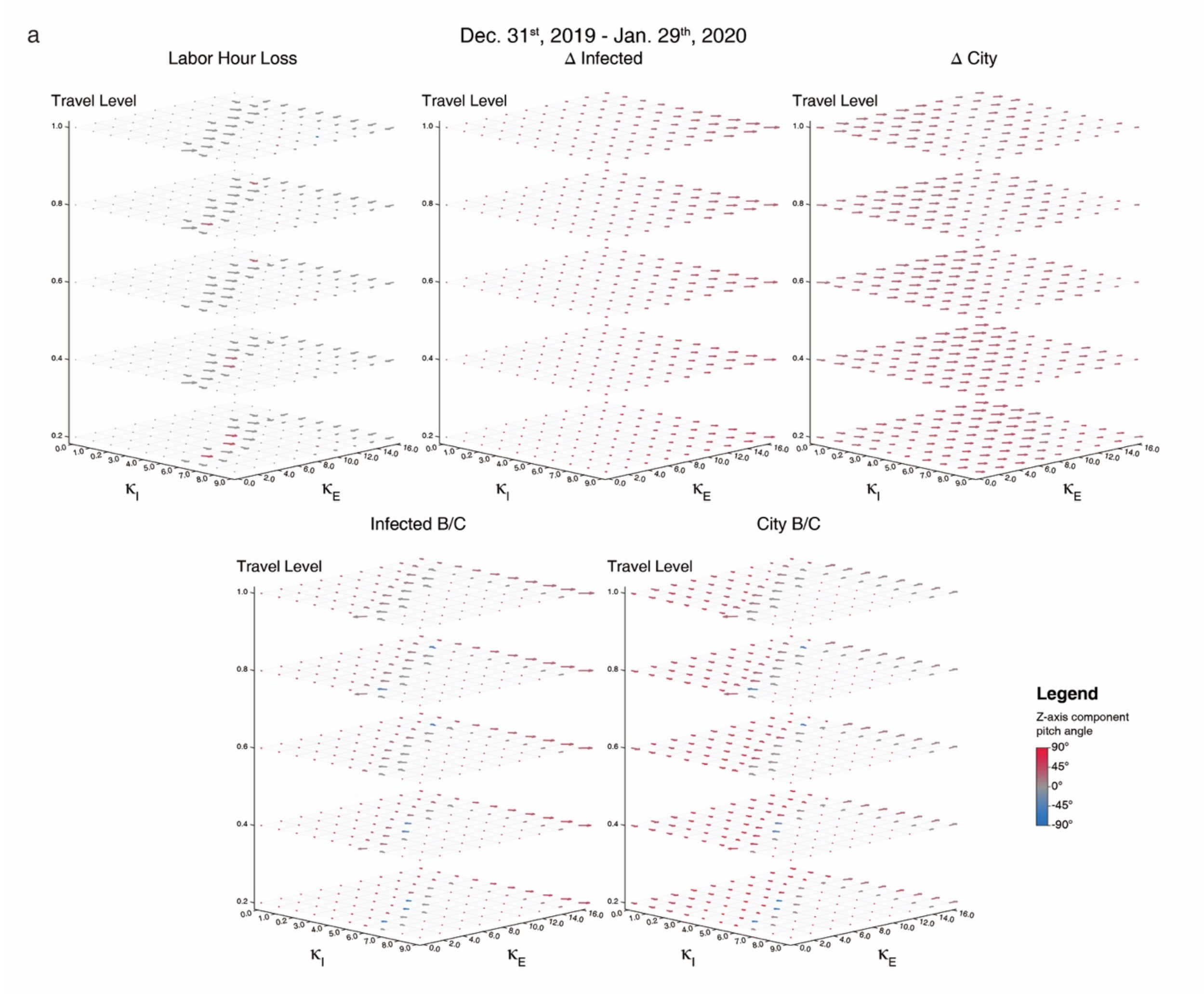

**Fig. S4. Global gradients of the cost, effectiveness, and cost-effectiveness functions at the accelerating sage.**

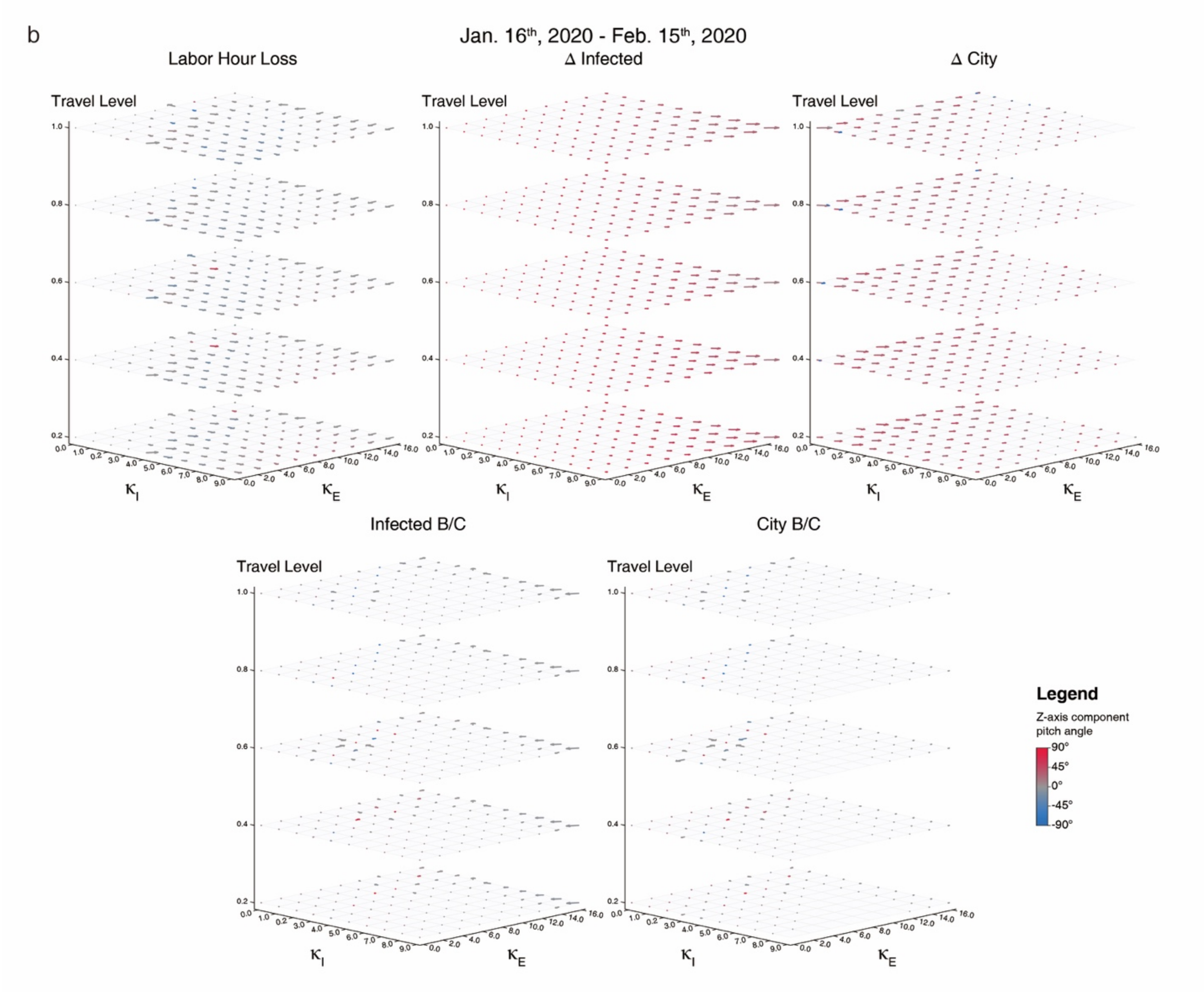

**Fig. S5. Global gradients of the cost, effectiveness, and cost-effectiveness functions at the peak stage.**

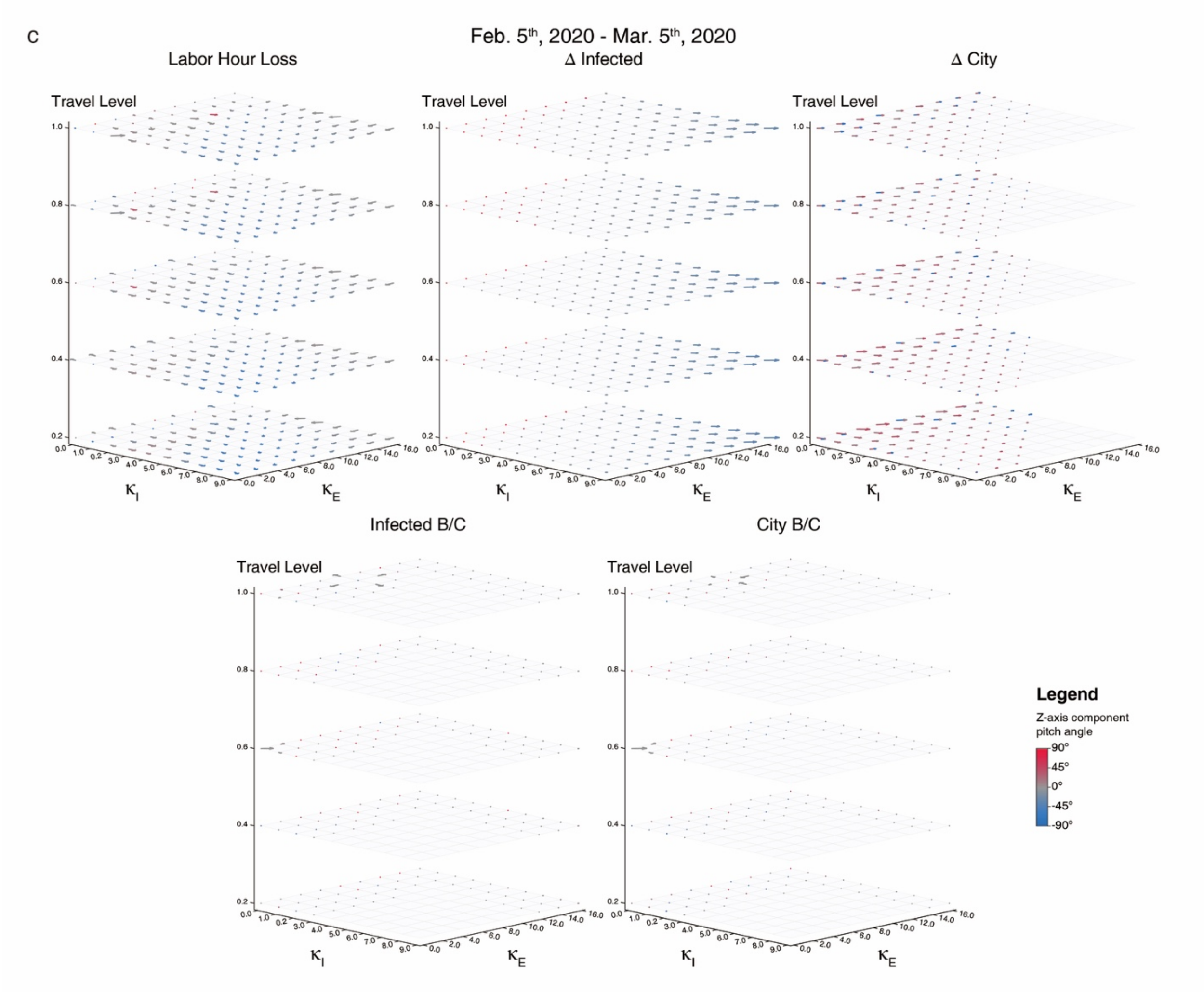

**Fig. S6. Global gradients of the cost, effectiveness, and cost-effectiveness functions at the declining stage.**

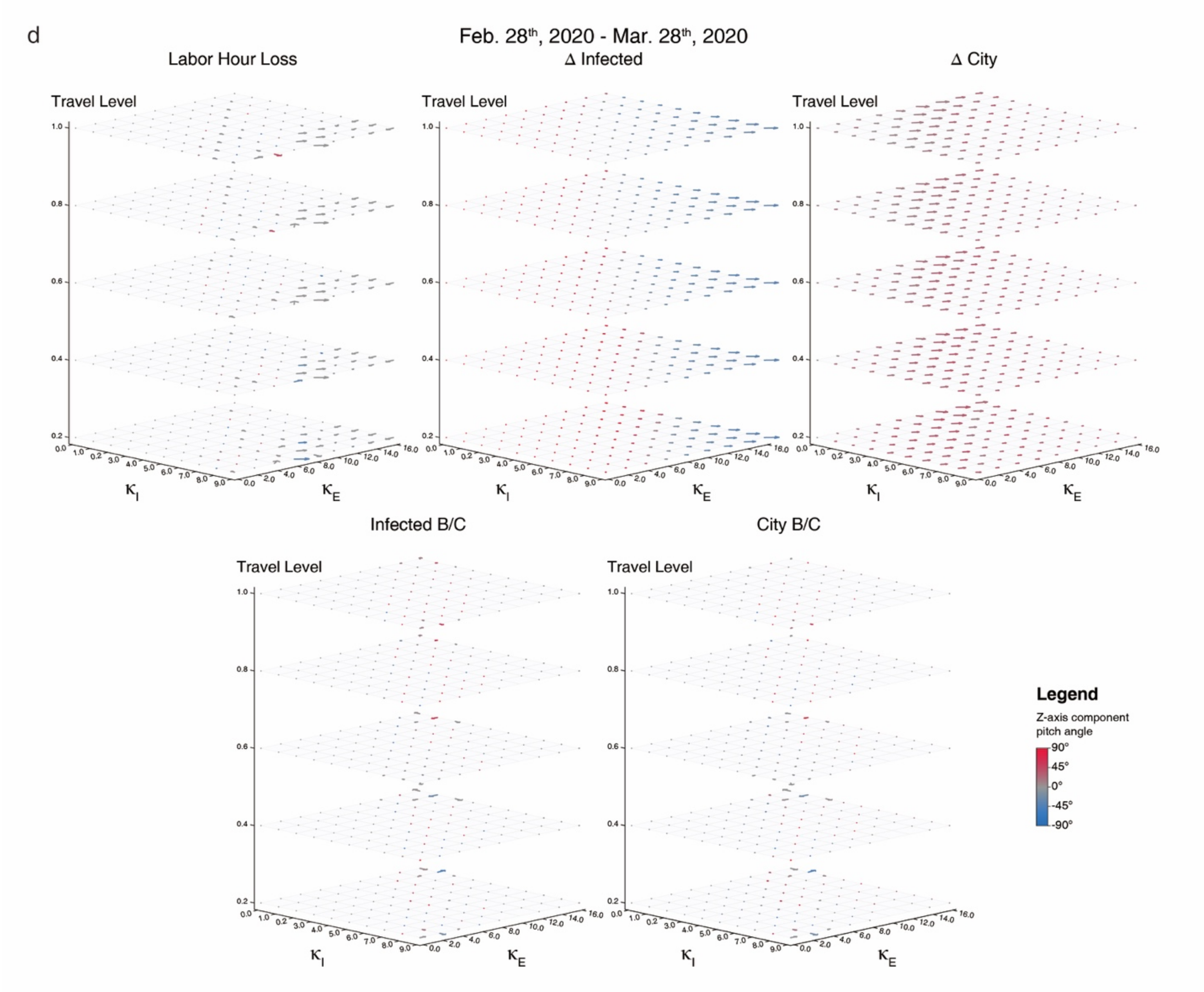

**Fig. S7. Global gradients of the cost, effectiveness, and cost-effectiveness functions at the ending stage.**

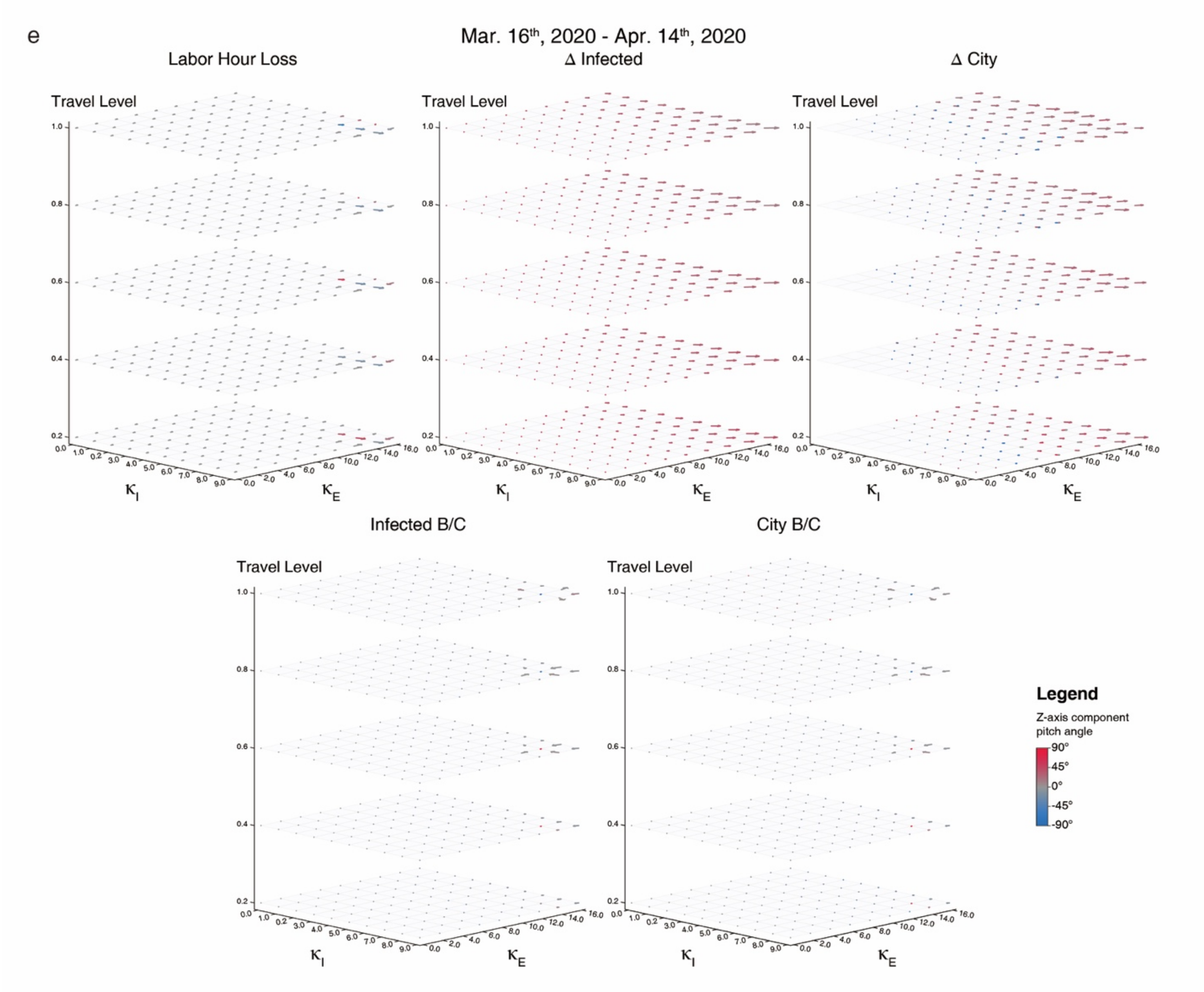

**Table S1. The relationship between control measures and parameters of the STEx-SEIR model.**

| | $\kappa_I$ | $\kappa_E$ | TL |
|---|---|---|---|
| Regular epidemiological control | ++ | + | N/A |
| In-city social interaction control | + | ++ | N/A |
| Intercity travel Restriction | N/A | N/A | + |

**Table S2. Threshold values for $\kappa_I$ and $\kappa_E$ of the "elimination" and "control" strategies.**

| | Elimination ($R_0$ << 1) | Control ($R_0$ < 1) |
|---|---|---|
| $\kappa_I$ | 1/3 of the normal value while $\kappa_E$ is the normal value | 2/3 of the normal value while $\kappa_E$ is the normal value |
| $\kappa_E$ | 1/4 of the normal value while $\kappa_I$ is the normal value | 1/2 of the normal value while $\kappa_I$ is the normal value |

**Table S3. Mortality rates based on the onset days at different stages.**

| Onset day | No. cumulative confirmed cases | No. deaths | Mortality rate (%) |
|---|---|---|---|
| Before December 31, 2019 | 104 | 15 | 14.4 |
| January 1–10, 2020 | 757 | 102 | 15.6 |
| January 11–20, 2020 | 6,174 | 310 | 5.7 |
| January 21–31, 2020 | 32,642 | 494 | 1.9 |
| February 1–11, 2020 | 44,672 | 102 | 0.8 |

Source: Epidemiology Working Group for NCIP Epidemic Response, 2020

**Table S4. Mortality rates in different regions at differernt stages.**

| Onset day | Wuhan | Hubei except Wuhan | China except Hubei |
|---|---|---|---|
| December 8–31, 2019 | 14.4 | 10.0% | 0% |
| January 1–10, 2020 | 17.2 | 11.9% | 5.0% |
| January 11–20, 2020 | 7.2 | 5.0% | 1.0% |
| January 21–31, 2020 | 2.4 | 1.7% | 0.3% |
| February 1–11, 2020 | 1.1 | 0.8% | 0.1% |

**Table S5. Initial values of $\kappa_I$ and $\kappa_E$ in different regions.**

| Region | $\kappa_I$ | $\kappa_E$ |
|---|---|---|
| Wuhan | 9.1 | 16.1 |
| Hubei province except Wuhan | 10.55 | 16.1 |
| All except Hubei province | 12.8 | 24 |

**Table S6. Summary of policies of 32 provinces and selected cities.**

| Policy category | Number of provinces (32 total) | Policy | Number of provinces (32 total) |
|---|---|---|---|
| Long-distance travel restrictions | 29 | Travel ban | 2 |
| | | Closing of highways | 7 |
| | | Closing of road passenger transport | 16 |
| | | Closing of trains and ferries | 1 |
| | | Reduction of train frequencies | 8 |
| | | Health screening at terminals | 5 |
| In-city travel restrictions | 9 | Shutdown of taxi services | 2 |
| | | Citywide travel ban | 5 |
| | | Shutdown of buses | 4 |
| | | Private car travel ban | 1 |
| | | Private car travel restrictions | 1 |
| Residential area control | 32 | Exit & entry screening | 4 |
| | | Lockdown | 15 |
| | | Limited exit & entry | 2 |
| | | Closing | 2 |
| Public space control | 20 | Essentials only | 12 |
| | | Cancelling public events | 11 |
| | | Enhanced sanitary enforcement | 3 |
| | | Requiring QR-code scanning at entry & exit for potential tracing purposes | 2 |
| Shutdown of workplaces | 32 | Until February 3 | 4 |
| | | Until February 10 | 25 |
| | | Until February 14 | 1 |
| | | Indefinitely | 2 |
| Closing of schools | 32 | Until February 17 | 8 |
| | | Until February 24 | 2 |
| | | Until early March | 14 |
| | | Indefinitely | 8 |

**Table S7. Gradients of $\kappa_I$, $\kappa_E$, and TL for the two effectiveness functions at the early stage of the epidemic.**

| | $\kappa_I$ | $\kappa_E$ | TL |
|---|---|---|---|
| ***Number of Infected Cases*** | | | |
| Gradient (1st decile) | 9.6% | 12% | 2.4% |
| Gradient (10th decile) | 0.41% | 0.14% | 0.12% |
| ***Number of Cities with Infected Cases*** | | | |
| Gradient (1st decile) | 1.3% | 1.9% | 0.91% |
| Gradient (10th decile) | 4.8% | 1.9% | 1.8% |

**Table S8. Gradients of $\kappa_I$, $\kappa_E$, and TL for the two effectiveness functions at the accelerating stage of the epidemic.**

| | $\kappa_I$ | $\kappa_E$ | TL |
|---|---|---|---|
| ***Number of Infected Cases*** | | | |
| Gradient (1st decile) | 12% | 14% | 0.21% |
| Gradient (10th decile) | 0.60% | 0.13% | 0.035% |
| ***Number of Cities with Infected Cases*** | | | |
| Gradient (1st decile) | 0.42% | 0.57% | 0 |
| Gradient (10th decile) | 4.0% | 0.79% | 0.68% |

**Table S9. Gradients of $\kappa_I$, $\kappa_E$, and TL for the two effectiveness functions at the peak stage of the epidemic.**

| | $\kappa_I$ | $\kappa_E$ | TL |
|---|---|---|---|
| ***Number of Infected Cases*** | | | |
| Gradient (1st decile) | 15% | 16% | -0.51% |
| Gradient (10th decile) | 0.66% | 0.059% | 0.016% |
| ***Number of Cities with Infected Cases*** | | | |
| Gradient (1st decile) | 0 | 0 | 0 |
| Gradient (10th decile) | 5.8% | 1.1% | 1.3% |

**Table S10. Gradients of $\kappa_I$, $\kappa_E$, and TL for the two effectiveness functions at the declining stage of the epidemic.**

| | $\kappa_I$ | $\kappa_E$ | TL |
|---|---|---|---|
| ***Number of Infected Cases*** | | | |
| Gradient (1st decile) | 16% | 23% | -0.80% |
| Gradient (10th decile) | 0.023% | 0.066% | 0.020% |
| ***Number of Cities with Infected Cases*** | | | |
| Gradient (1st decile) | 0.91% | 1.5% | 0.12% |
| Gradient (10th decile) | 2.88% | 0.60% | 0 |

**Table S11. Gradients of $\kappa_I$, $\kappa_E$, and TL for the two effectiveness functions at the ending stage of the epidemic.**

| | $\kappa_I$ | $\kappa_E$ | TL |
|---|---|---|---|
| ***Number of Infected Cases*** | | | |
| Gradient (1st decile) | 11% | 26% | 0.88% |
| Gradient (10th decile) | 0.094% | 0.19% | 0.063% |
| ***Number of Cities with Infected Cases*** | | | |
| Gradient (1st decile) | 9.4% | 35% | 3.2% |
| Gradient (10th decile) | 0 | 0 | 0 |